\documentclass[aps,prb,twocolumn,superscriptaddress,showpacs]{revtex4}

\usepackage{graphicx}
\usepackage{amssymb}
\usepackage{amsmath}
\usepackage{latexsym}
\usepackage{ulem}
\usepackage{color}
\usepackage{dcolumn}
\usepackage[T1]{fontenc}
\usepackage[utf8]{inputenc}
\usepackage[english]{babel}
\usepackage[colorlinks=true,linkcolor=blue,citecolor=blue]{hyperref}%
\usepackage[caption=false]{subfig}	
\usepackage[load-configurations = abbreviations]{siunitx}

\begin{document}

\title{Boosting the power factor with resonant states: a model study}

\author{S. Th\'ebaud}
\email[E-mail:]{simon.thebaud@univ-lyon1.fr}
\affiliation{Univ Lyon, Universit\'e Claude Bernard Lyon 1, CNRS, Institut Lumi\`ere Mati\`ere, F-69622, LYON, France}
\author{Ch. Adessi}
\affiliation{Univ Lyon, Universit\'e Claude Bernard Lyon 1, CNRS, Institut Lumi\`ere Mati\`ere, F-69622, LYON, France}
\author{S. Pailh\`es}
\affiliation{Univ Lyon, Universit\'e Claude Bernard Lyon 1, CNRS, Institut Lumi\`ere Mati\`ere, F-69622, LYON, France}
\author{G. Bouzerar}
\affiliation{Univ Lyon, Universit\'e Claude Bernard Lyon 1, CNRS, Institut Lumi\`ere Mati\`ere, F-69622, LYON, France}

\begin{abstract}
A particularly promising pathway to enhance the efficiency of thermoelectric materials lies in the use of resonant states, as suggested by experimentalists and theorists alike. In this paper, we go over the mechanisms used in the literature to explain how resonant levels affect the thermoelectric properties, and we suggest that the effects of hybridization are crucial yet ill-understood. In order to get a good grasp of the physical picture and to draw guidelines for thermoelectric enhancement, we use a tight-binding model containing a conduction band hybridized with a flat band. We find that the conductivity is suppressed in a wide energy range near the resonance, but that the Seebeck coefficient can be boosted for strong enough hybridization, thus allowing for a significant increase of the power factor. The Seebeck coefficient can also display a sign change as the Fermi level crosses the resonance. Our results suggest that in order to boost the power factor, the hybridization strength must not be too low, the resonant level must not be too close to the conduction (or valence) band edge, and the Fermi level must be located around, but not inside, the resonant peak.
\end{abstract}

\pacs{}
\maketitle

\section{Introduction}

The prospect of an unprecedented global crisis, due to our energy dependance toward unclean fossile ressources such as oil and coal, has led to a tremendous push in research for clean and sustainable energy sources. Consequently, there has been a renewed interest toward thermoelectric materials in the field of condensed matter physics. \cite{Zheng_renewable_energy_applications, Elsheikh_renewable_energy} Thermoelectric generators that could be competitive on an industrial scale require \cite{Goldsmid_Introduction_to_Thermoelectricity, Rowe_CRC_Handbook_of_Thermoelectrics, Zhang_thermoelectric_material_energy_conversion, Zheng_renewable_energy_applications} thermoelectric materials exhibiting a large figure of merit ${zT}$, defined as 
\[
{zT} = \frac{\sigma S^2}{\kappa} T,
\]
in which $\sigma$ is the electrical conductivity, $S$ the Seebeck coefficient, $\kappa$ the thermal conductivity, and $T$ the temperature. The higher ${zT}$ is, the closer the device efficiency is to the Carnot efficiency. In the past decades, the drive to develop efficient thermoelectric devices has led scientists to search for green and industry-friendly materials exhibiting a figure of merit of at least 1. The standard procedure for boosting ${zT}$ has been to decrease the thermal conductivity by using alloyed, nanostructured, or nanocomposite materials. \cite{Kanatzidis_nanostructured_new_paradigm, Dresselhaus_bulk_nanostructured_materials, Vineis_big_gains_small_features, Snyder_complex_thermoelectric_materials} However, there is a growing concern that this method will soon reach its limits and that further significant progress will be achieved by focusing on the power factor $\sigma S^2$.

Hicks and Dresselhaus breathed new life into the field when they theorized that the electronic part of the figure of merit could be greatly enhanced by spatially confining known thermoelectric materials in two-dimensional quantum wells \cite{Hicks_Dresselhaus_quantum_well_structure_figure_of_merit} or in one-dimensional quantum wires. \cite{Hicks_Dresselhaus_1D} These papers gave rise to a great drive towards the synthesis of materials exhbiting a reduction of dimensionality, \cite{Pichanusakorn_nanostructured_thermoelectrics, Heremans_low_dimensional_thermoelectricity, Dresselhaus_new_directions_low_dimensional} by trapping electrons either in superlattices, for instance PbTe/Pb$_{1-x}$Eu$_x$Te \cite{Hicks_experimental_effect_quantum_well} and Bi$_2$Te$_3$/Sb$_2$Te$_3$, \cite{Venkatasubramanian_thin_film} or in the interface between two compounds, such as TiO$_2$ and SrTiO$_3$. \cite{Ohta_giant_seebeck_2D_electron_gas_STO} The basic idea is to increase the density of state (DOS) near the band edge in order to boost the power factor. A few years later, Mahan and Sofo\cite{Mahan_Sofo_best_thermoelectric} suggested a mechanism by which the introduction of a sharp peak in the DOS could significantly boost the figure of merit. This can be achieved by doping a bulk compound with a dopant that creates sharp impurity states inside and hybridized with the conduction or valence band of the host material: such states are called resonant states or virtual bound states. A great virtue of this method is its applicability to both thin films and bulk materials, unlike electronic confinement which usually requires thin films. The procedure has been attempted by many researchers (see the review by Heremans et al.\cite{Heremans_resonant_levels_semiconductor}), for instance in the case of Thallium doped PbTe, \cite{Heremans_Enhancement_of_Thermoelectric_Efficiency_in_PbTe, Nemov_Ravish_review_thallium_lead_chalcogenides} Tin doped Bi$_2$Te$_3$, \cite{Jaworski_Heremans_resonant_level_tin} Indium doped SnTe \cite{Zhang_resonant_Indium_SnTe, Tan_high_performance_SnTe_resonant_levels} or Aluminium doped PbSe. \cite{Zhang_enhancement_resonant_states_PbSe} It has also been suggested \cite{Mahan_Sofo_best_thermoelectric} that the very high power factor of YbAl$_3$ \cite{Li_Sn_substitution_YbAl3, Rowe_transport_properties_YbAl3} could be explained by the presence of a sharp f-level peak in its DOS. \cite{Zhou_electronic_structure_YbAl3} On the theoretical side, the calculation of thermoelectric transport properties is most often achieved by using first principle density functional theory simulations and the Boltzmann equation under the constant relaxation time approximation, methods that are material-specific and often quite demanding in computation time.   \cite{Yang_enhancing_thermoelectric_doping_electronegativites_distinct, Singh_thermopower_PbTe_boltzmann, Peng_electronic_structure_transport_properties_doped_PbSe, Parker_Singh_high_temperature_dependance_heavily_doped_PbSe, Yang_electronic_properties_CdTe_doped} 

Despite such intense activity around resonant states in thermoelectricity, there are still a few widespread misconceptions about how resonant levels affect thermoelectric transport. In this paper, our aim is to put forward a clear physical picture and provide a few guidelines to best achieve a resonant enhancement of the power factor. In the second part, we will discuss the mechanisms as presented in most of the existing literature, and point out their limitations. In a third part, we will present an alternative physical picture with the help of a simple tight-binding model. This model will allow us to identify some important trends and the role of the main physical parameters, independantly of any material-specific band structure or scattering law. 

\section{The Transport Distribution Function}

The electronic transport properties of any compound depend on a single material-dependant function of the energy and temperature, called the transport distribution function (TDF) $\Sigma(E,T)$. Once the TDF is known, the electrical conductivity $\sigma$ and the Seebeck coefficient $S$ at a given temperature $T$ and temperature-dependant chemical potential $\mu(T)$ can be obtained as \cite{ashcroft}
\begin{equation}
\sigma (T) = \int dE \left( - \frac{\partial f}{\partial E} \right) \Sigma(E,T) , 
\end{equation}
\begin{equation}
S (T) = -\frac{1}{e T \sigma}\int dE \left( - \frac{\partial f}{\partial E} \right) \left(E-\mu\right) \Sigma(E,T) ,
\end{equation}
in which $e$ is the elementary charge and $f$ the Fermi distribution. Since $\left( - \frac{\partial f}{\partial E} \right) $ is basically a window function of width $\approx 4 k_B T$ centered on $\mu$, these relations simplify at low temperature to
\begin{equation}
\sigma (T) \approx \Sigma(E=\mu,T) ,
\end{equation}
\begin{equation}
S (T) \approx -\frac{\pi^2 k_B^2 T}{3 e} \frac{1}{ \Sigma}\frac{\partial \Sigma}{\partial E}{(E=\mu,T)} ,
\end{equation}
which are the well-known Mott formula. \cite{Rowe_CRC_Handbook_of_Thermoelectrics,ashcroft} Therefore, to maximize the power factor $\sigma S^2$, one needs to find a material in which the TDF has a high amplitude (high conductivity) as well as sharp variations (high Seebeck coefficient). 

\begin{figure}[h!]
\subfloat{\includegraphics[width=0.49\columnwidth]{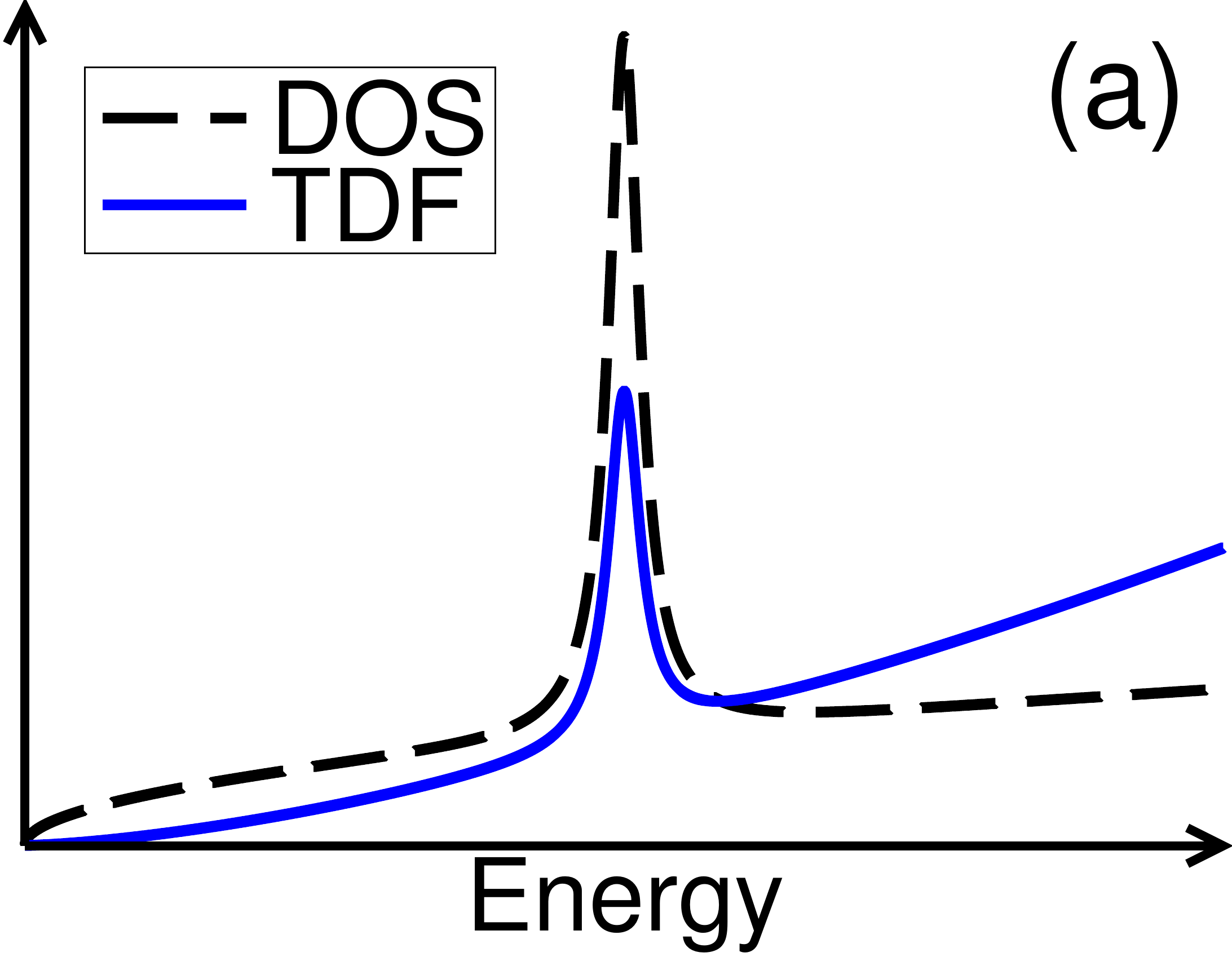}}
\vspace{0cm}
\subfloat{\includegraphics[width=0.49\columnwidth]{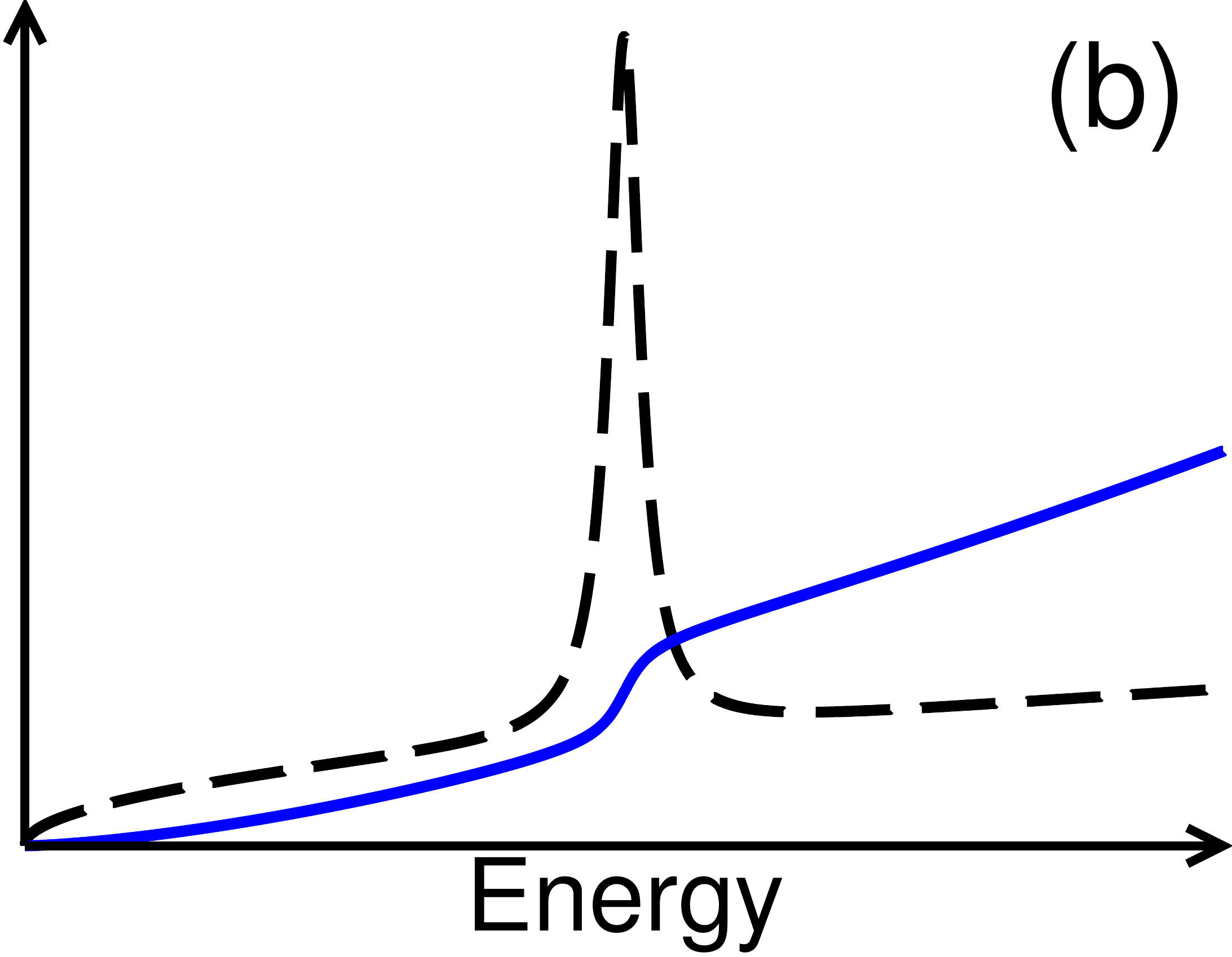}}
\caption{(Color online) Schematic representation of the density of state (dashed black line) and transport distribution function (blue line) in the presence of a resonant peak, in the Mahan-Sofo picture (a) and in the "carrier density" picture (b).}
\label{fig1}
\end{figure}

Mahan and Sofo \cite{Mahan_Sofo_best_thermoelectric} found that the shape of the TDF most conducive to good thermoelectric properties is the delta function. Writing the Einstein formula \cite{ashcroft}
\begin{equation}
\Sigma(E,T) = e^2 \, g(E) \, v_x^2(E) \, \tau(E,T),
\end{equation}
in which $g(E)$ is the DOS, $v_x(E)$ is the band velocity along the transport direction $x$, and $\tau(E,T)$ is the relaxation time, they suggested that a sharp peak in the DOS would translate to a sharp peak in the TDF (Fig.~\ref{fig1}(a)), provided that the velocity $v_x$ is not suppressed significantly around resonance (as we will see, the velocity is actually strongly affected in most cases, the picture of Fig.~\ref{fig1}(a) is therefore usually invalid). In this case, they also calculated that the presence of a background DOS could destroy the resonant effect, which would encourage looking for resonant states very near the host valence or conduction band edge. Notice how the sign of the slope of the TDF changes around the resonant state in this physical picture. Consequently, one should expect the sign of the Seebeck coefficient to change as the Fermi level crosses the resonant band. This interpretation has been invoked to explain the results of ab-initio calculations for ZnSe containing O impurities. \cite{Lee_enhancing_thermoelectric_mismatched_doping}

Another slightly different mechanism is used to explain the effects of resonant states throughout the literature. \cite{Heremans_resonant_levels_semiconductor, Yang_theory_experiment_perspective, Heremans_Enhancement_of_Thermoelectric_Efficiency_in_PbTe, Zhang_enhancement_resonant_states_PbSe} Since the TDF is basically the conductivity at low temperature, one can write  
\begin{equation}
\Sigma (E,T) \approx e \, n(E) \, \mu_x(E,T) ,
\end{equation}
in which $n(E)$ is the total carrier density and $\mu_x(E,T)$ is the carrier mobility along the transport direction. The Mott formula then yields
\begin{equation}
\label{seebeck_heremans}
S \approx -\frac{\pi^2 k_B^2 T}{3 e} \left( \frac{1}{n}\frac{\partial n}{\partial E} + \frac{1}{\mu_x}\frac{\partial \mu_x}{\partial E} \right) . 
\end{equation}
The first term, $\frac{1}{n}\frac{\partial n}{\partial E} = \frac{g(E)}{n(E)}$, depends very little on temperature and is associated with band structure effects, it is straighforward to see that a local increase in the DOS would cause a local boost in the Seebeck coefficient. The second term, $\frac{1}{\mu_x}\frac{\partial \mu_x}{\partial E}$, is associated in the literature with scattering. If the resonant states bring about a drastic change in the carrier scattering, the Seebeck coefficient might be boosted through this term. The effects of resonant scattering are thought to vanish at high temperature, as the scattering becomes dominated by phonons. Therefore, this mechanism is probably unable to help in designing materials for power generation (which usually involves high temperatures), and is not the topic of this paper which is focused on band structure effects. Assuming no resonant scattering, then, the carrier mobility is thought not to change violently aroung the resonance, and the TDF is approximatively proportional to the carrier density $n$ (Fig.~\ref{fig1}(b)). Note that in this picture, the slope of the TDF remains positive, so the sign of the Seebeck coefficient is not expected to change. Moreover, the Fermi level should be located right at the resonance for maximum effect. 

Both mechanisms in Fig.~\ref{fig1} require the resonant states to participate in conduction, meaning that either the band velocity $v_x$ or the mobility $\mu_x$ must not be drastically suppressed by the resonance. This seems incompatible with the requirement that the resonant peak in the DOS be sharp since a flat band implies low velocities and low mobilities. The mobility is indeed not simply a scattering quantity: one can see from the Drude formula $\mu_x = e \tau /m_x^*$  that a large effective band mass $m_x^*$ suppresses the mobility, and a flat band is precisely composed of heavy carrier states (consequently, associating the term $\frac{1}{\mu_x}\frac{\partial \mu_x}{\partial E}$ of equation~\ref{seebeck_heremans} to resonant scattering only and discarding it on this ground is incorrect in many cases). Thus, a single very flat band, for instance a doping impurity band in the gap of the host material, leads to bad thermoelectric properties despite having a very delta-like DOS. \cite{Zhou_Dresselhaus_optimal_bandwidth_efficiency_thermoelectrics, Jeong_best_bandstructure_thermoelectric_landauer} Confinement techniques work around this problem because they introduce flat dispersions only in the confinement directions. For the same reason, highly anisotropic orbitals like d orbitals are naturally conducive to good thermoelectric properties, and it is possible to boost the power factor by making them come into play. \cite{Bilc_low_dimensional_band_engineering_highly_directional} Even then, a recent study \cite{Bilc_first_principles_modeling_sto} clearly shows that they must have both a strong weight in the DOS and a sufficient dispersion in the transport direction to enhance the thermoelectric properties. Another technique, called band convergence, \cite{Pei_band_engineering} consists in engineering a material to add other valleys at the bottom of the conduction band or at the top of the valence band, this has been achieved for instance in Selenium doped PbTe. \cite{Pei_Snyder_convergence_electronic_bands_high_performance} The sharpness of the DOS increase is then abandoned, and the DOS is, in effect, multiplied by a factor: at fixed Fermi energy, the conductivity is increased while the Seebeck coefficient is unchanged. However, in the case of a sharp peak in the DOS of a bulk material, which is the focus of this paper, there are two types of carrier states: light carriers, corresponding to the undistorted bands of the material, and heavy carriers, corresponding to the peak in the DOS. As noted by Heremans et al. \cite{Heremans_resonant_levels_semiconductor}, if the two types of states are not hybridized with one another, the transport properties are entirely dominated by the light carriers (which have the highest conductivity) and the sharp peak is useless. Therefore, some hybridization is needed to boost the power factor.
 
This raises a few questions: What exactly happens to the TDF when there is hybridization? What is the right hybridization strength? Where should the resonant level be, with respect to the edge of the host band? Where should the Fermi level be, with respect to the resonant level? What kind of boost can we expect for the power factor? In order to answer these questions and investigate the typical behaviour of the relevant quantities independantly of the material-specific band-structure or scattering law, we shall make use of a tight-binding model containing a flat band hybridized with a conduction band. Contrary to the parabolic band model, a tight-binding description let us explore the entire band structure, and thus choose freely the position of the Fermi level and the resonant band. Such tight-binding models have been shown to successfully describe the electronic structure of oxides in confinement problems \cite{Zhong_Held_quantum_confinement_oxide_tight_binding, Zhong_Held_microscopic_understanding_orbital_splitting_oxide_interfaces} and the bulk thermoelectric properties of SrTiO$_3$. \cite{Bouzerar_unified_modelling_thermoelectric_properties_STO}

\section{Tight-binding model}

In order to study the typical effects of resonant states on the thermoelectric properties in the simplest and clearest way, we write a spinless tight-binding Hamiltonian on a simple cubic lattice ($N = L^3$ sites, lattice constant $a$) featuring an extended conduction band and a heavy resonant band:
\begin{equation}
{\cal H} = {\cal H}_0 + {\cal H}_{V} ,
\end{equation}
in which
\begin{equation}
{\cal H}_0 = -t_0 \sum_{\bf{<i,j>}} \left( c^{\dagger}_{\bf{i}} c_{\bf{j}} + c^{\dagger}_{\bf{j}} c_{\bf{i}} \right) + E_l \sum_{\bf{i}} l^{\dagger}_{\bf{i}} l_{\bf{i}} ,
\end{equation}
and ${\cal H}_{V}$ is the hybridization term:
\begin{equation}
{\cal H}_{V} = V \sum_{\bf{i}} \left( c^{\dagger}_{\bf{i}} l_{\bf{i}} + l^{\dagger}_{\bf{i}} c_{\bf{i}} \right) ,
\end{equation}
with $<\bf{i,j}>$ the pairs of nearest neighboring sites in the lattice, $c_{\bf{i}}$ the electron annihilation operator on the site $\bf{i}$ in an extended conduction orbital, and $l_{\bf{i}}$ the electron annihilation operator on the site $\bf{i}$ in a heavy resonant orbital. As the spin degree of freedom is irrelevant in this study, we write a spinless Hamiltonian for the sake of clarity, and will simply take into account the spin degeneracy by introducing a factor 2 in our numerical results.

The first term in ${\cal H}_0$ gives rise to a conduction band (extended states) of width $12 t_0$ and dispersion $\epsilon({\bf k}) = -2 t_0 (\cos(k_x a) + \cos(k_y a) + \cos(k_z a))$, $t_0$ being the hopping constant. As there is no hopping term for the resonant orbitals, those give rise to a flat band (heavy states): the second term in ${\cal H}_0$ governs the position $E_l$ of this flat band in the DOS. ${\cal H}_{V}$ hybridizes the two orbitals (extended and heavy) in each lattice site, with a hybridization strength $V$. If $V = 0$, there is no hybridization and the heavy states do not participate in conduction at all.   

Given the simplicity of this model, we do not aim in this paper to reproduce experimental results. Rather, our goal is to investigate the general consequences of hybridizing a flat band with an extended conduction band, and to clarify the role of each physical parameters. Thus, we are to compare the thermoelectric transport properties in the absence of resonance (which corresponds to the reference case $V=0$) to the same properties in the presence of hybridization (resonant case).

\begin{figure}[h!]
\includegraphics[width=1.0\columnwidth]{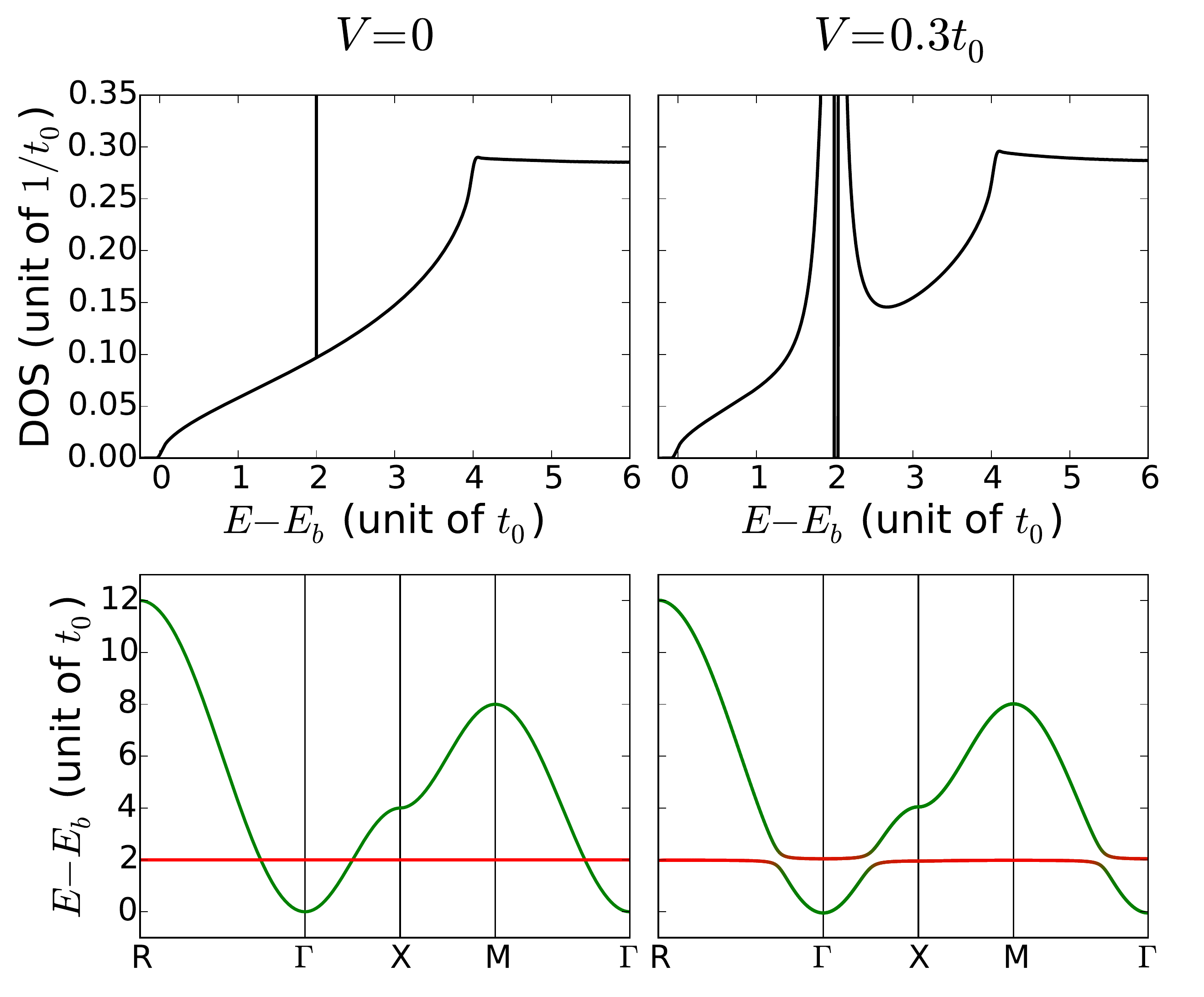}
\caption{(Color online) DOS and dispersions for $V = 0$ (left) and $V = 0.3 t_0$ (right) with $E_l - E_b = 2 t_0$. $E_b$ designates the bottom of the lower band. The green color denotes extended conduction orbital character and red denotes heavy resonant orbital character. Different values of $V$ yield qualitatively similar results.}
\label{fig2}
\end{figure}

The Hamiltonian ${\cal H}$ can be diagonalized straightforwardly. The lower half of the DOS and the two-band dispersions are plotted in Fig.~\ref{fig2}, for $V = 0$ and $V = 0.3 t_0$, with $E_l - E_b = 2t_0$ ($E_b$ being the bottom of the conduction band). Several values for $V$ and $E_l$ will be subsequently considered. If the orbitals are hybridized, the peak in the DOS broadens and, in this model, a hybridization gap appears. 

To separate the influence of scattering in electronic transport from the effects of the band structure, we write the TDF as
\begin{equation}
\Sigma(E,T) = D(E) \, \tau(E,T) ,
\end{equation}
in which $D(E)$ is the Drude weight. \cite{Scalapino_insulator_metal_criteria, Kohn_theory_insulating_state} The Drude weight is computed by inserting a magnetic flux $\phi$ along the transport direction and taking the derivative of the many-body ground state energy, $E_0$:
\begin{equation}
D = \frac{1}{a L} \frac{\partial^2 E_0}{\partial \phi^2} .
\end{equation}
It also satisfies the sum rule \cite{Millis_Coppersmith_optical_spectral_weight, Bouzerar_optical_conductivity} 
\begin{equation}
\label{sumrule}
D + \frac{2}{\pi} \int_0^{\infty} d \omega \, \sigma_{reg}(\omega) = \frac{e^2 t_0}{a \hbar^2} \frac{\left< -\hat{K_x} / N \right> }{t_0}
\end{equation}
in which $\sigma_{reg}(\omega)$ is the regular part of the optical conductivity and $\hat{K_x}$ is, in this model, the kinetic energy along the transport direction $x$. $D_0 = \frac{e^2 t_0}{a \hbar^2}$ will henceforth be the unit of Drude weight. For our translationally invariant Hamiltonian, the regular conductivity is associated to interband transitions and we checked that the sum rule is satisfied. As for the relaxation time, a realistic expression for $\tau$ would be highly material-dependant (dominating scattering process, presence of disorder and impurities, etc...) and remember that the effects of resonant scattering are supposed to disappear at high temperature. This paper investigates the consequences on thermoelectric transport of distorsions in the TDF caused through band structure effects by the presence of a resonant band. Therefore, we use the constant relaxation time approximation: $\tau = \tau_0$, with $\tau_0$ a constant. Consequently, the TDF is simply proportional to the Drude weight, which makes $D(E)$ the central quantity of this study. However, it is important to note that using a constant relaxation time is not always justified. Further studies investigating the effects of resonant scattering (which is beyond the scope of this manuscript) are needed to go above this first approximation.  

\begin{figure}[h!]
\includegraphics[width=1.0\columnwidth]{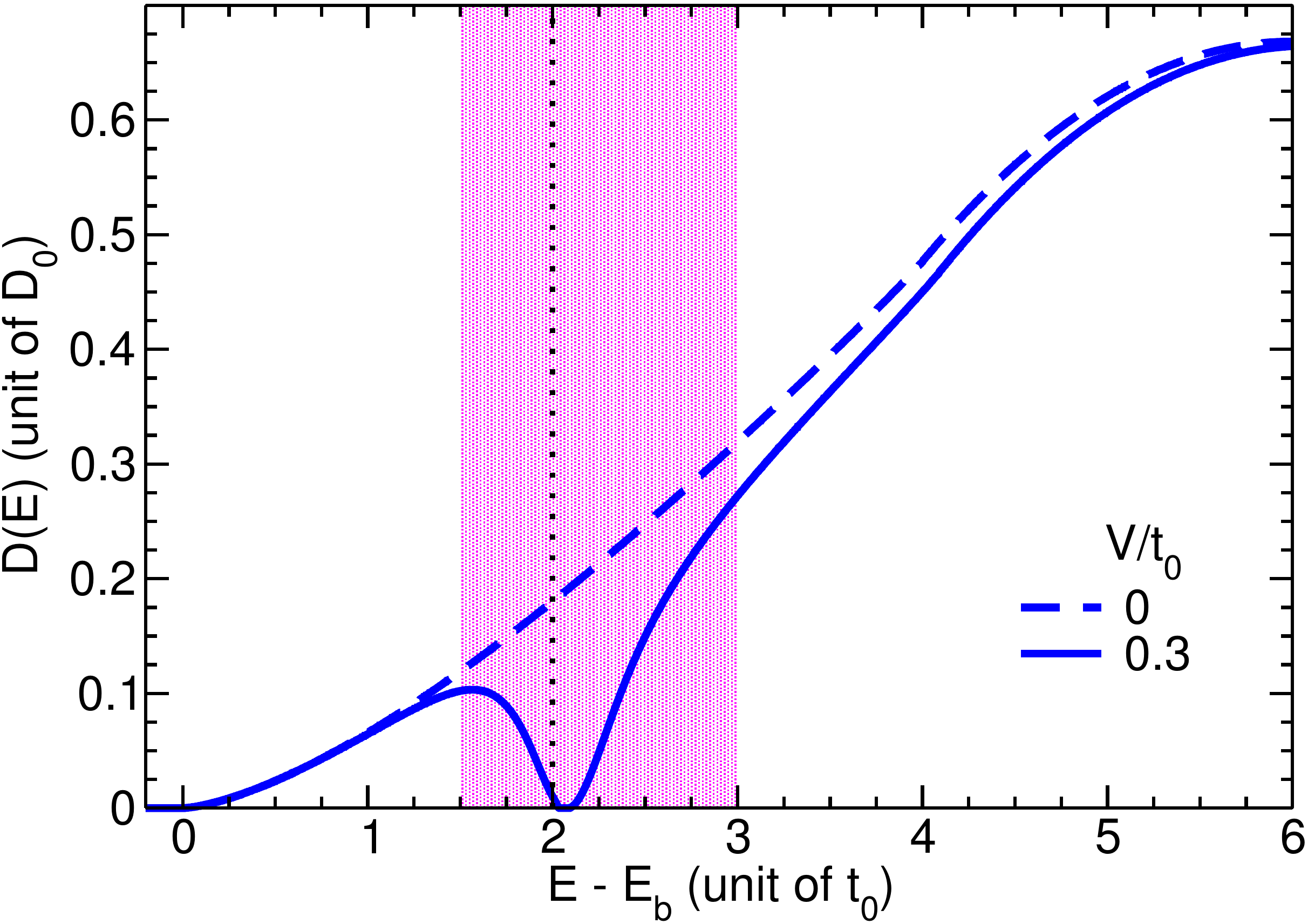}
\caption{(Color online) Drude weight in unit of $D_0 = \frac{e^2 t_0}{a \hbar^2}$ for $V=0$ (dashed line) and for $V = 0.3 t_0$ (solid line) with $E_l - E_b = 2 t_0$ (a dotted vertical line marks the position of the flat band). Different values of $V$ yield qualitatively similar results. The shading marks the zone in which the Drude weight is suppressed.}
\label{fig3}
\end{figure}

The Drude weight as a function of the Fermi energy is shown in Fig.~\ref{fig3} for the reference ($V = 0$) and the resonant case ($V = 0.3 t_0$, $E_l - E_b = 2 t_0$), in unit of $D_0$. For the reference case, the regular conductivity vanishes and the Drude weight identifies with the right-hand side of equation~\ref{sumrule}, which is a rather smooth and monotonic function of the energy. In contrast, in the resonant case, there is a substantial drop in the Drude weight within the resonant band region. This is due to the mixing of the heavy and light carriers near $E_l$, the latter acquire a heavy character which drastically suppress their conduction. It is important to compare Fig.~\ref{fig3} to the mechanisms represented in Fig.~\ref{fig1}, notice how the effects of resonant states on the TDF are unlike anything proposed in the literature. Drawing the consequences for the thermoelectric properties, this result appears at first to be unfavorable. The TDF, and consequently the electrical conductivity, is strongly suppressed within a large energy region around the resonance (shaded area on Fig.~\ref{fig3}), this does not bode well indeed for the power factor. There is, however, a silver lining: by creating a dip in the Drude weight, the resonant band also introduces steep variations. Therefore, we expect the Seebeck coefficient to be boosted around the resonance and, since it is squared in the power factor, we can still hope that this effect will compensate the loss in conductivity and enhance the power factor. 

In order to find out if this is true or not, we now present some results for the transport coefficients, giving special attention to comparisons between the power factor with and without resonance.

\begin{figure}[h!]
\subfloat{\includegraphics[width=1.0\columnwidth]{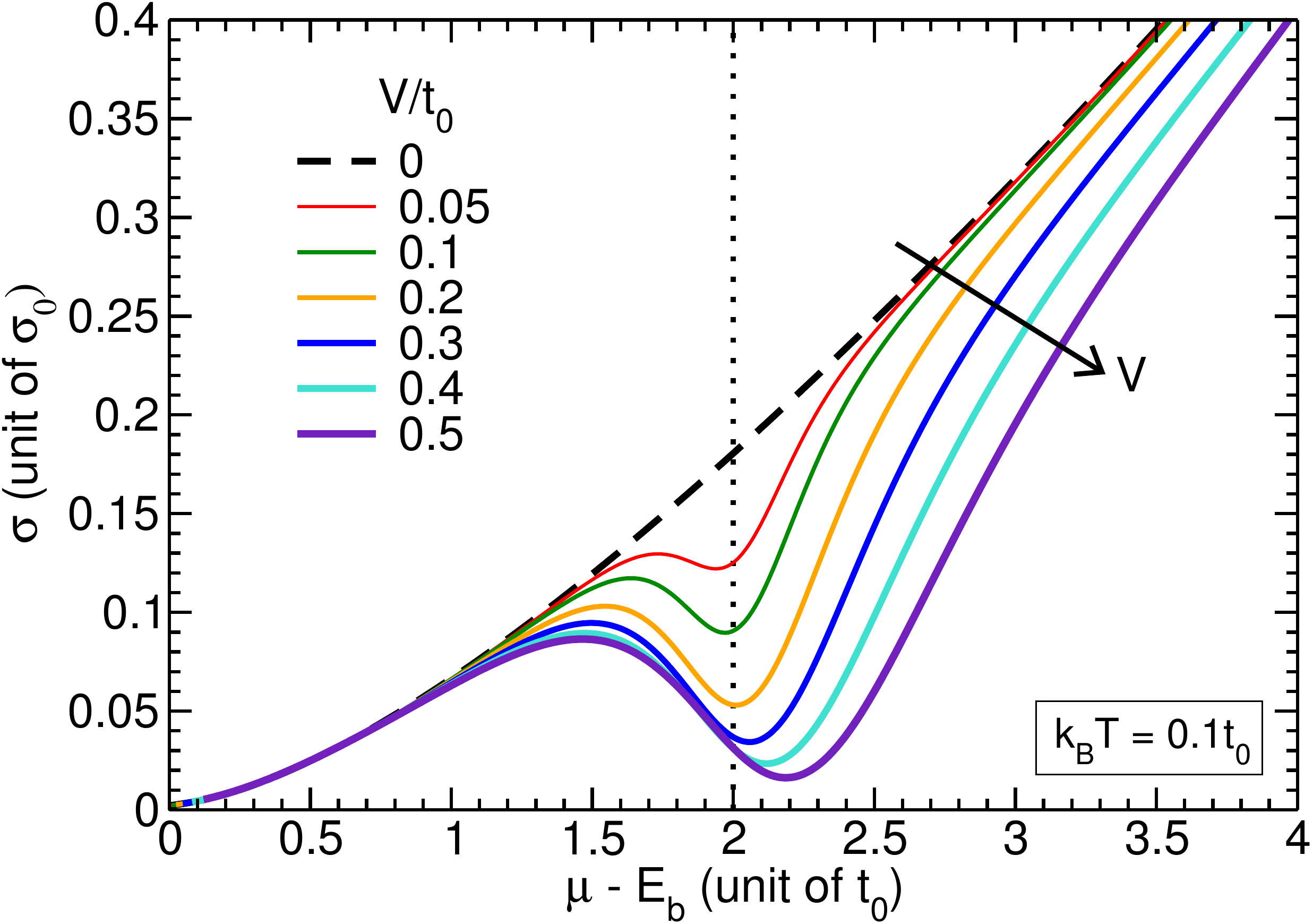}}
\hspace{0cm}
\subfloat{\includegraphics[width=1.0\columnwidth]{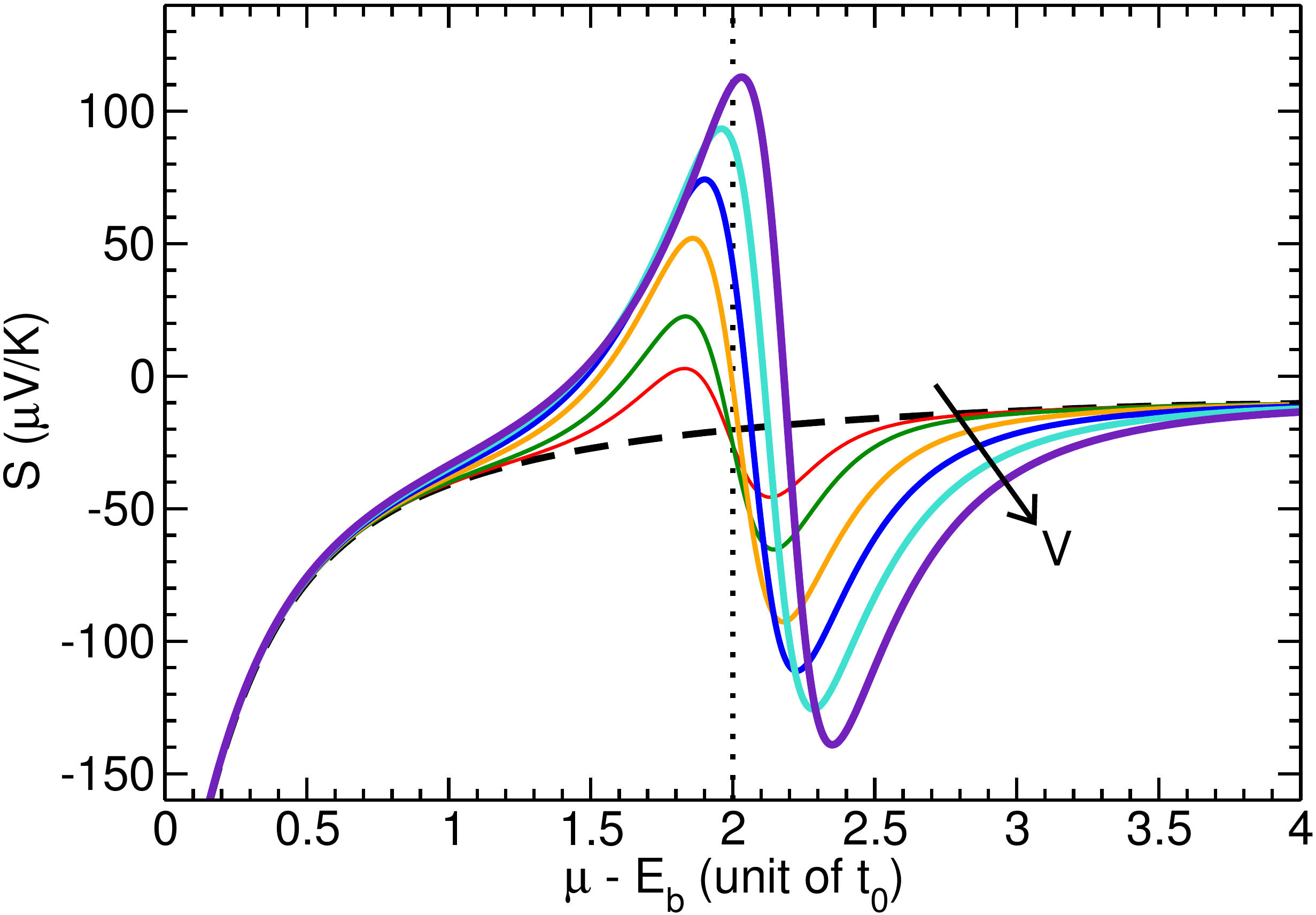}}
\caption{(Color online) Top: electrical conductivity in unit of $\sigma_0 = \frac{e^2}{a \hbar} \frac{t_0 \tau_0}{\hbar}$ against $\mu$ for various $V$, $E_l - E_b = 2 t_0$ and $k_B T = 0.1 t_0$ (approximately room temperature). The reference case ($V=0$) is the dashed black line. The flat band position is marked by a dotted black line. Bottom: Seebeck coefficient in $\mu V/K$.}
\label{sigma_V}
\end{figure}

Setting the resonant band position at $E_l - E_b = 2 t_0$, we investigate the effect of the hybridization amplitude $V$. We show in Fig.~\ref{sigma_V} the electrical conductivity (in unit of $\sigma_0 = \frac{e^2}{a \hbar} \frac{t_0 \tau_0}{\hbar}$) and the Seebeck coefficient (in $\mu V/K$) against the chemical potential $\mu$, for several values of $V$. The temperature is set at $k_B T = 0.1 t_0$, which corresponds to room temperature for $t_0 \approx \SI{0.25}{\electronvolt}$ (this is
typically the case in SrTiO$_3$ \cite{Bouzerar_unified_modelling_thermoelectric_properties_STO} for instance). The resonant band position is marked by a dotted black line. As $V$ is turned on from the reference case (dashed black line), the conductivity is suppressed near the resonant band. As for the Seebeck coefficient, there is a boost on each side of the resonance and a change of sign for high enough values of $V$. The sign change is very interesting, because we are still in the conduction band. This might prove extremely useful for materials in which n-type doping is possible but p-type is difficult, or vice versa. The boost in Seebeck coefficient is typically comparable or greater than the decrease in electrical conductivity, so we expect the power factor (in which the Seebeck coefficient appears squared) to get boosted.

\begin{figure}[h!]
\includegraphics[width=1.0\columnwidth]{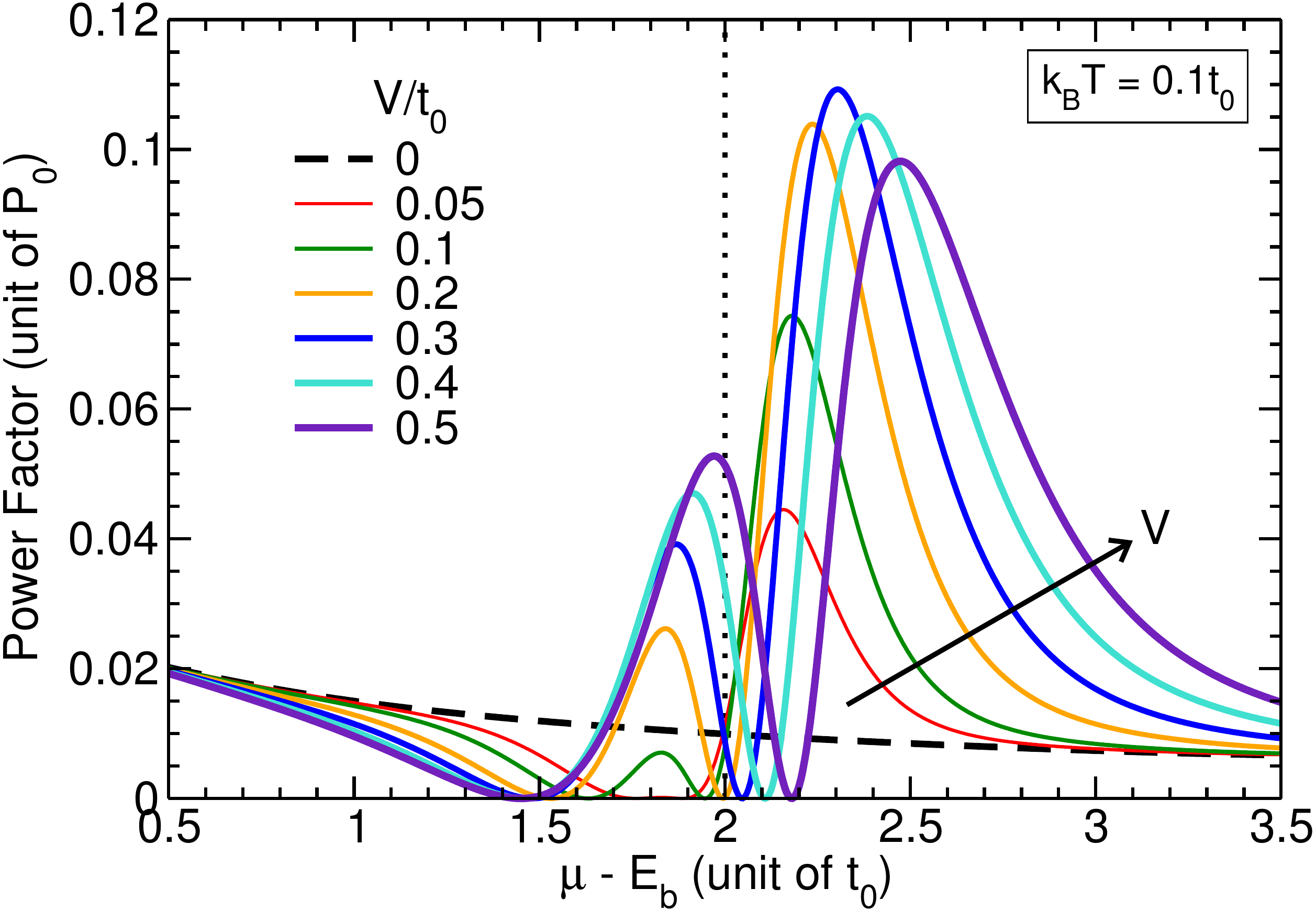}
\caption{(Color online) The power factor in unit of $P_0 = \frac{k_B^2}{a \hbar} \frac{t_0 \tau_0}{\hbar}$ against $\mu$ for various $V$, $E_l - E_b = 2 t_0$, and $k_B T = 0.1 t_0$ (approximately room temperature). The reference case ($V=0$) is the dashed black line. The flat band position is marked by a dotted black line.}
\label{pf_V}
\end{figure}

The power factor is plotted in Fig.~\ref{pf_V} in units of $P_0 = \frac{k_B^2}{a \hbar} \frac{t_0 \tau_0}{\hbar}$ for the same values of $V$, $E_l$, and $T$. There is a boost above and below the resonant band for high enough values of the hybridization, which confirms that hybridization plays an central role in resonant enhancement of the thermoelectric properties. Also, notice that the power factor is suppressed when the Fermi level is located inside the resonant band, unlike the mechanism in which the TDF is proportional to the carrier density. There appears to be an optimal value of $V$ around $0.3 t_0$, for which the power factor is an order of magnitude higher than its reference value at $\mu - E_b \approx 2.3 t_0$. As this optimal value is not very dependant on the resonant band position, we set $V = 0.3 t_0$ to discuss the influence of the other parameters, namely the resonant band position $E_l$ and the temperature $T$. 

\begin{figure}[ht!]
\includegraphics[width=1.0\columnwidth]{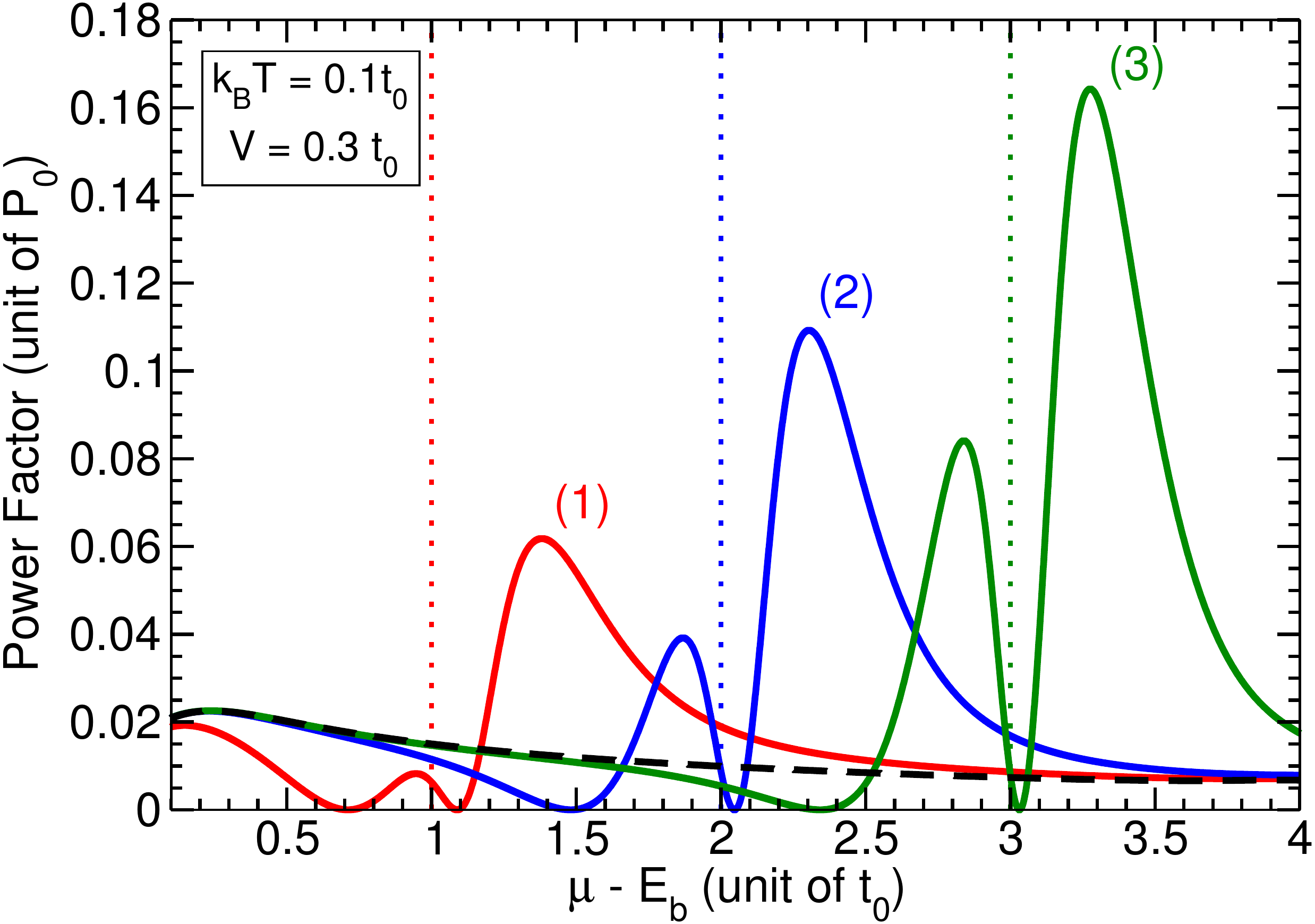}
\caption{(Color online) The power factor in unit of $P_0 = \frac{k_B^2}{a \hbar} \frac{t_0 \tau_0}{\hbar}$ against $\mu$ for $V = 0.3 t_0$, $k_B T = 0.1 t_0$ (approximately room temperature) and $E_l - E_b = t_0$ (1), $E_l - E_b = 2 t_0$ (2), $E_l - E_b = 3 t_0$ (3). The flat band position in each case is marked by a dotted line. The reference case ($V=0$) is the dashed black line.}
\label{pf_eps}
\end{figure}

Turning our attention to the influence of the resonant band position $E_l$, we show the power factor against $\mu$ in Fig.~\ref{pf_eps} for $k_B T = 0.1 t_0$ and three different band positions ($E_l - E_b = t_0$, $2 t_0$, and $3 t_0$, marked by dotted vertical lines), along with the reference (dashed black line). It is manifest that the closest the resonant band is to the center of the conduction band, the more boosted the power factor is. This makes sense: close to the center, the Drude weight is higher (see Fig.~\ref{fig3}) and so the drop created by the resonant band is steeper. In fact, if the flat band lies outside the conduction band (below $E_b$), then the transport properties are the same as the reference case. Therefore, to achieve a significant boost of the power factor, the resonant band must be located well within the conduction band. This runs counter to the advice of Mahan and Sofo \cite{Mahan_Sofo_best_thermoelectric} to have a background DOS as small as possible. Of course, a resonant band located too far from the edge of the conduction band would be inaccessible with reasonable doping, so the optimal location for the resonant level might be the result of a compromise.

\begin{figure}[h!]
\subfloat{\includegraphics[width=0.99\columnwidth]{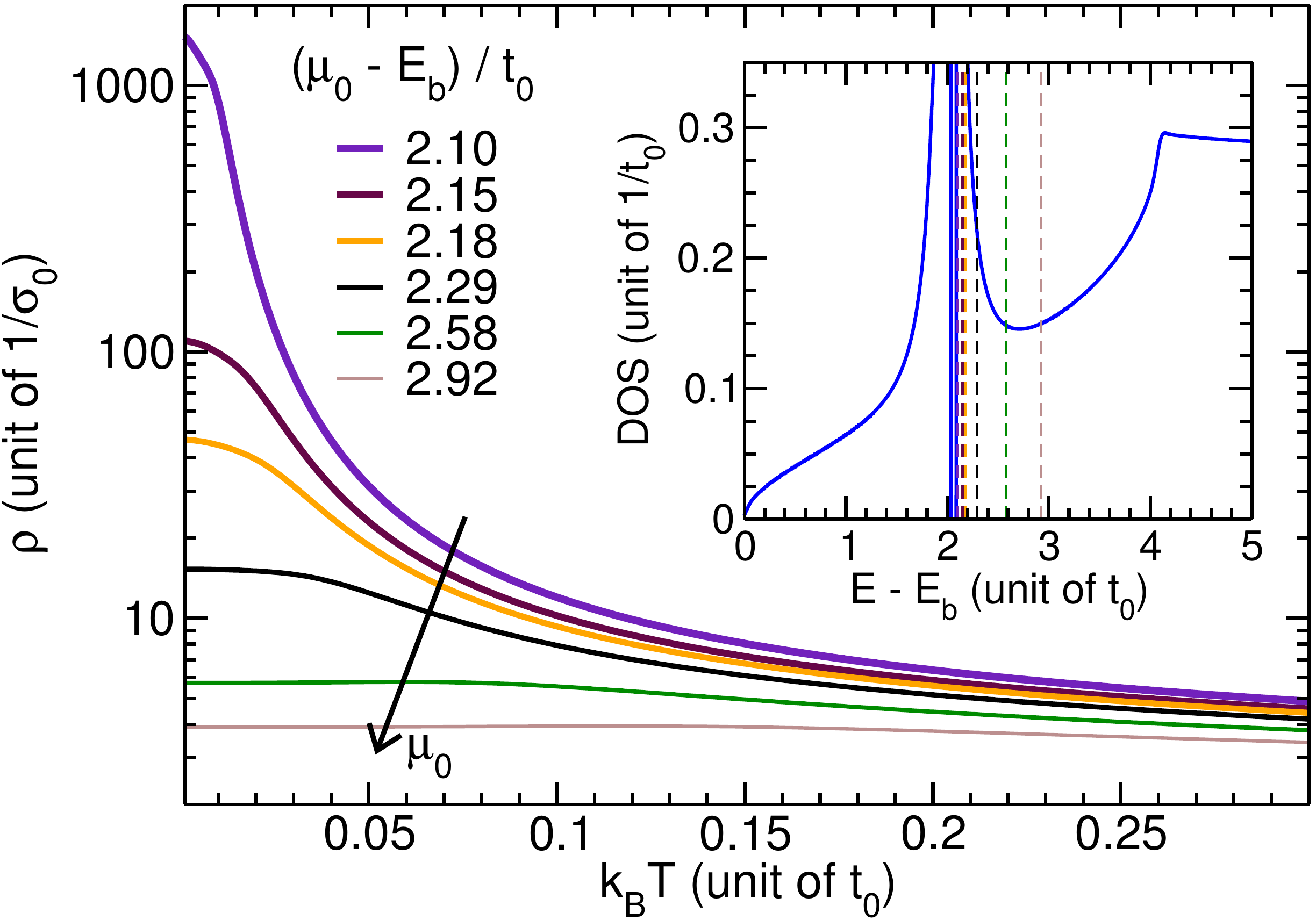}}
\hspace{0cm}
\subfloat{\includegraphics[width=0.99\columnwidth]{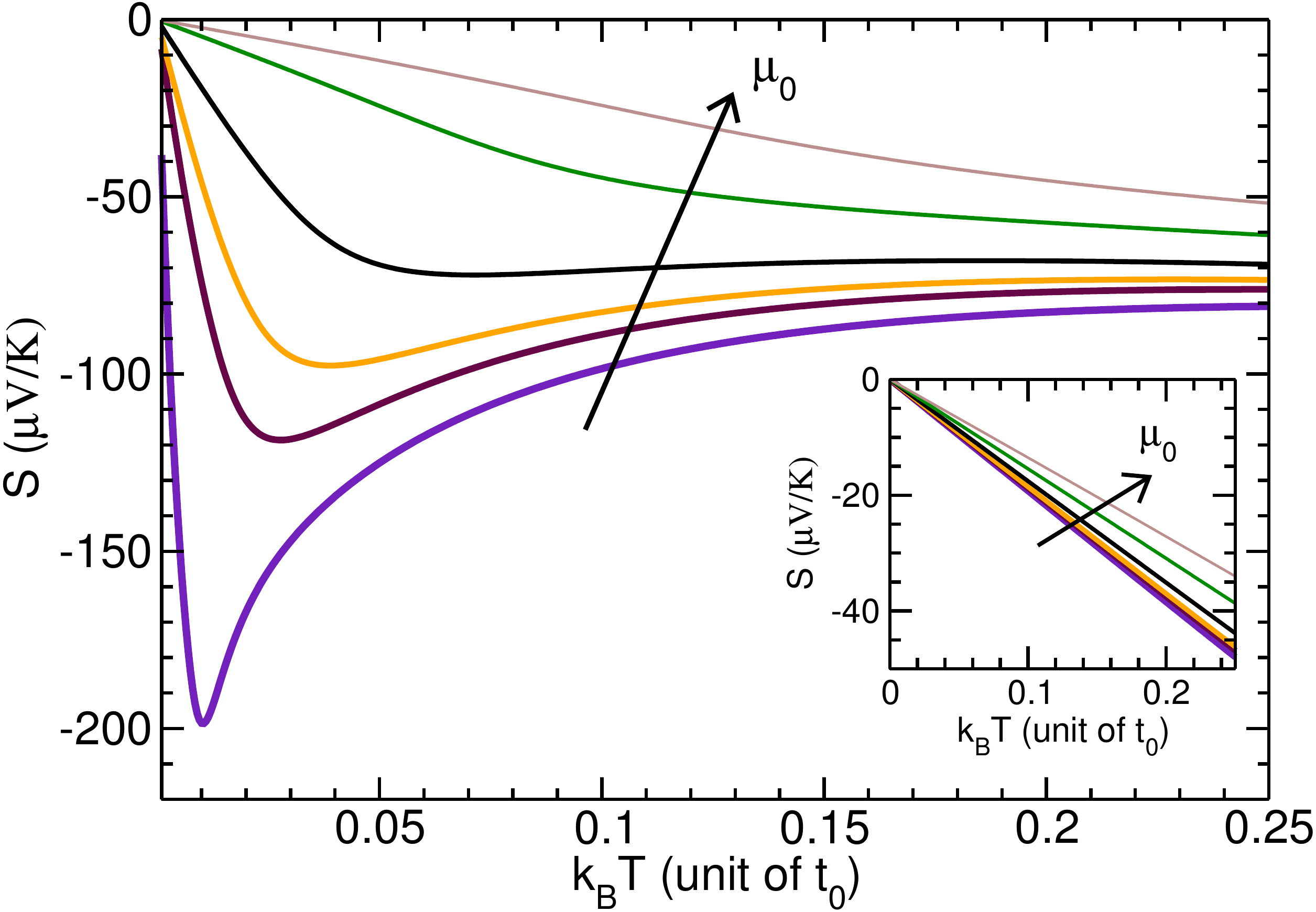}}
\caption{(Color online) Top: the resistivity in unit of $\frac{1}{\sigma_0} = \frac{a \hbar}{e^2} \frac{\hbar}{t_0 \tau_0}$ and logarithmic scale against the temperature for $V = 0.3 t_0$, $E_l - E_b = 2 t_0$, and six different electron densities. Each density is labeled by the chemical potential at $T = \SI{0}{\kelvin}$, $\mu_0$, shown in inset with the DOS. Bottom: the Seebeck coefficient in $\mu V/K$ against the temperature, with the reference case shown in inset for the same $\mu_0$. Notice the different scales for the Seebeck coefficient.}
\label{rho_T}
\end{figure}

We show in Fig.~\ref{rho_T} the resistivity $\rho$ (in log scale and units of $1/\sigma_0$) and the Seebeck coefficient against the temperature for six electron densities, each labeled by the chemical potential at $T = \SI{0}{\kelvin}$, $\mu_0$, shown in inset of the resistivity. As the temperature changes, the electron density is fixed but the chemical potential varies. The reference case is shown in inset for the Seebeck coefficient. Due to the heavy nature of the states near the resonance, the resistivity displays a metal-insulator transition as $\mu_0$ draws closer to the resonant level. It is informative to study the behaviour of the Seebeck coefficient as this metal-insulator transition takes place. Comparing the reference and resonant cases, the Seebeck coefficient is not only boosted but also displays a very different and non-standard behaviour, exhibiting a minimum at various temperatures if $\mu_0$ is close enough to the resonance. It is important to keep in mind that such a structure may emerge from the distorsion of the Drude weight caused by the presence of a resonant band.

\begin{figure}[h!]
\includegraphics[width=1.0\columnwidth]{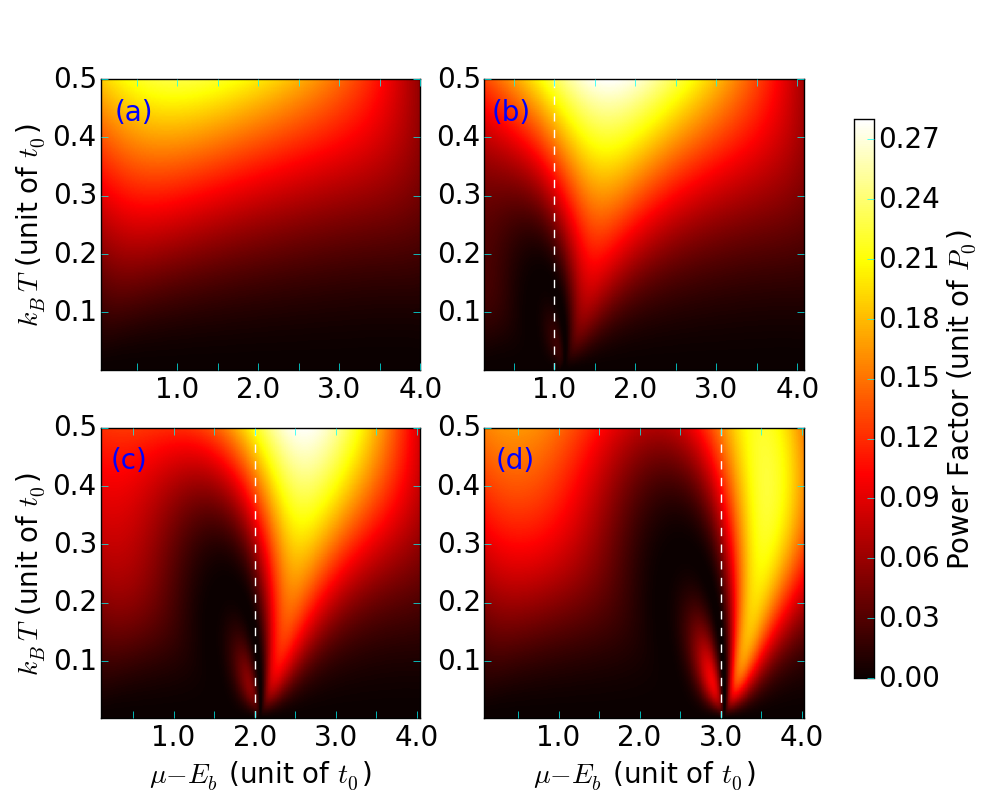}
\caption{(Color online) Color map of the power factor in unit of $P_0$ on the $(\mu,T)$ plane for the reference case $V=0$ (a) and the resonant case $V = 0.3 t_0$ with $E_l-E_b = t_0$ (b), $E_l-E_b = 2 t_0$ (c), and $E_l-E_b = 3 t_0$ (d). The dashed white lines mark the position of the flat band.}
\label{colorgraph}
\end{figure}

In order to get an overview of the power factor with respect to the chemical potential and the temperature, we plot in Fig.~\ref{colorgraph} a color map of the power factor on the $(\mu,T)$ plane in the reference case $V = 0$ (a) and the resonant case $V = 0.3 t_0$ with $E_l-E_b = t_0$ (b), $E_l-E_b = 2 t_0$ (c), and $E_l-E_b = 3 t_0$ (d). From panel (a), significantly increasing the power factor in the reference case requires going to temperatures as high as $\SI{1200}{\kelvin}$ (assuming $t_0 \approx \SI{0.25}{\electronvolt}$). However, if a resonant band is present (panels (b), (c) and (d)), the region of high power factor is dragged down to much lower temperatures, a process that is more efficient the closer the resonant band is to the center of the conduction band.

In summary, we showed that it is possible to boost the power factor by introducing resonant states. We went over the mechanisms found in the literature about how resonant levels can enhance thermoelectric transport. To clarify the physical picture, the crucial role of hybridization, and the influence of the relevant physical parameters, we used a minimal tight-binding model featuring an extended conduction band hybridized with a heavy flat band. In contrast to the scenarios usually proposed, the presence of the resonant band was shown to introduce a significant dip in the Drude weight or, equivalently, in the transport distribution function, and thus to suppress the conductivity. However, the strong variations of the transport distribution function lead to a boost in the Seebeck coefficient and overcompensates the conductivity reduction. This competition allows for a large increase in the power factor. We determined that a significant boost requires that the hybridization is high enough (typically 30$\%$ of the hopping), that the resonant band is not located too close to the conduction band edge, and that the Fermi level is located near but not inside the resonant band. This new understanding of the mechanism by which resonant levels may influence thermoelectric transport opens the door to more sophisticated studies investigating for instance the effects of the density of resonant states, the anisotropy of the orbitals, and the nature of the hybridization (on-site or not, with multiple atoms...). As our tight-binding model is compatible with the description of more complicated band structures and can easily accomodate the presence of disorder, it constitutes the basis from which such further studies can be carried out, hopefully leading to a better understanding of resonant doping.

\begin{footnotesize}

\end{footnotesize}


\begin{thebibliography}{49}
\expandafter\ifx\csname natexlab\endcsname\relax\def\natexlab#1{#1}\fi
\expandafter\ifx\csname bibnamefont\endcsname\relax
  \def\bibnamefont#1{#1}\fi
\expandafter\ifx\csname bibfnamefont\endcsname\relax
  \def\bibfnamefont#1{#1}\fi
\expandafter\ifx\csname citenamefont\endcsname\relax
  \def\citenamefont#1{#1}\fi
\expandafter\ifx\csname url\endcsname\relax
  \def\url#1{\texttt{#1}}\fi
\expandafter\ifx\csname urlprefix\endcsname\relax\def\urlprefix{URL }\fi
\providecommand{\bibinfo}[2]{#2}
\providecommand{\eprint}[2][]{\url{#2}}

\bibitem[{\citenamefont{Zheng et~al.}(2011)\citenamefont{Zheng, Liu, and
  Wang}}]{Zheng_renewable_energy_applications}
\bibinfo{author}{\bibfnamefont{X.}~\bibnamefont{Zheng}},
  \bibinfo{author}{\bibfnamefont{C.}~\bibnamefont{Liu}}, \bibnamefont{and}
  \bibinfo{author}{\bibfnamefont{Q.}~\bibnamefont{Wang}},
  \bibinfo{journal}{Renewable and Sustainable Energy Rev.}
  \textbf{\bibinfo{volume}{32}}, \bibinfo{pages}{486} (\bibinfo{year}{2011}).

\bibitem[{\citenamefont{ohamed Hamid~Elsheikh et~al.}(2014)\citenamefont{ohamed
  Hamid~Elsheikh, Shnawah, Sabri, Said, Hassan, Bashir, and
  Mohamad}}]{Elsheikh_renewable_energy}
\bibinfo{author}{\bibnamefont{ohamed Hamid~Elsheikh}},
  \bibinfo{author}{\bibfnamefont{D.~A.} \bibnamefont{Shnawah}},
  \bibinfo{author}{\bibfnamefont{M.~F.~M.} \bibnamefont{Sabri}},
  \bibinfo{author}{\bibfnamefont{S.~B.~M.} \bibnamefont{Said}},
  \bibinfo{author}{\bibfnamefont{M.~H.} \bibnamefont{Hassan}},
  \bibinfo{author}{\bibfnamefont{M.~B.~A.} \bibnamefont{Bashir}},
  \bibnamefont{and} \bibinfo{author}{\bibfnamefont{M.}~\bibnamefont{Mohamad}},
  \bibinfo{journal}{Renewable and Sustainable Energy Rev.}
  \textbf{\bibinfo{volume}{30}}, \bibinfo{pages}{337} (\bibinfo{year}{2014}).

\bibitem[{\citenamefont{Goldsmid}(2010)}]{Goldsmid_Introduction_to_Thermoelectricity}
\bibinfo{author}{\bibfnamefont{H.~J.} \bibnamefont{Goldsmid}},
  \emph{\bibinfo{title}{Introduction to Thermoelectricity}}
  (\bibinfo{publisher}{Springer}, \bibinfo{year}{2010}).

\bibitem[{\citenamefont{Rowe}(1995)}]{Rowe_CRC_Handbook_of_Thermoelectrics}
\bibinfo{author}{\bibfnamefont{D.}~\bibnamefont{Rowe}},
  \emph{\bibinfo{title}{CRC Handbooks of Thermoelectricity}}
  (\bibinfo{publisher}{CRC Press}, \bibinfo{year}{1995}).

\bibitem[{\citenamefont{Zhang and
  Zhao}(2015)}]{Zhang_thermoelectric_material_energy_conversion}
\bibinfo{author}{\bibfnamefont{X.}~\bibnamefont{Zhang}} \bibnamefont{and}
  \bibinfo{author}{\bibfnamefont{L.-D.} \bibnamefont{Zhao}},
  \bibinfo{journal}{J. Materiomics} \textbf{\bibinfo{volume}{1}},
  \bibinfo{pages}{92–105} (\bibinfo{year}{2015}).

\bibitem[{\citenamefont{Kanatzidis}(2010)}]{Kanatzidis_nanostructured_new_paradigm}
\bibinfo{author}{\bibfnamefont{M.~G.} \bibnamefont{Kanatzidis}},
  \bibinfo{journal}{Chem. Mater.} \textbf{\bibinfo{volume}{22}},
  \bibinfo{pages}{648–659} (\bibinfo{year}{2010}).

\bibitem[{\citenamefont{Minnich et~al.}(2009)\citenamefont{Minnich,
  Dresselhaus, F., and G.}}]{Dresselhaus_bulk_nanostructured_materials}
\bibinfo{author}{\bibfnamefont{A.~J.} \bibnamefont{Minnich}},
  \bibinfo{author}{\bibfnamefont{M.~S.} \bibnamefont{Dresselhaus}},
  \bibinfo{author}{\bibfnamefont{R.~Z.} \bibnamefont{F.}}, \bibnamefont{and}
  \bibinfo{author}{\bibfnamefont{C.}~\bibnamefont{G.}},
  \bibinfo{journal}{Energy Environ. Sci.} \textbf{\bibinfo{volume}{2}},
  \bibinfo{pages}{466} (\bibinfo{year}{2009}).

\bibitem[{\citenamefont{Vineis et~al.}(2010)\citenamefont{Vineis, Shakouri,
  Majumdar, and Kanatzidis}}]{Vineis_big_gains_small_features}
\bibinfo{author}{\bibfnamefont{C.~J.} \bibnamefont{Vineis}},
  \bibinfo{author}{\bibfnamefont{A.}~\bibnamefont{Shakouri}},
  \bibinfo{author}{\bibfnamefont{A.}~\bibnamefont{Majumdar}}, \bibnamefont{and}
  \bibinfo{author}{\bibfnamefont{M.~G.} \bibnamefont{Kanatzidis}},
  \bibinfo{journal}{Adv. Mater.} \textbf{\bibinfo{volume}{22}},
  \bibinfo{pages}{3970} (\bibinfo{year}{2010}).

\bibitem[{\citenamefont{Snyder and
  S.}(2008)}]{Snyder_complex_thermoelectric_materials}
\bibinfo{author}{\bibfnamefont{G.~J.} \bibnamefont{Snyder}} \bibnamefont{and}
  \bibinfo{author}{\bibfnamefont{T.~E.} \bibnamefont{S.}},
  \bibinfo{journal}{Nature. Mater.} \textbf{\bibinfo{volume}{7}},
  \bibinfo{pages}{105} (\bibinfo{year}{2008}).

\bibitem[{\citenamefont{Hicks and
  Dresselhaus}(1993{\natexlab{a}})}]{Hicks_Dresselhaus_quantum_well_structure_figure_of_merit}
\bibinfo{author}{\bibfnamefont{L.~D.} \bibnamefont{Hicks}} \bibnamefont{and}
  \bibinfo{author}{\bibfnamefont{M.~S.} \bibnamefont{Dresselhaus}},
  \bibinfo{journal}{Phys. Rev. B} \textbf{\bibinfo{volume}{47}},
  \bibinfo{pages}{12727} (\bibinfo{year}{1993}{\natexlab{a}}).

\bibitem[{\citenamefont{Hicks and
  Dresselhaus}(1993{\natexlab{b}})}]{Hicks_Dresselhaus_1D}
\bibinfo{author}{\bibfnamefont{L.~D.} \bibnamefont{Hicks}} \bibnamefont{and}
  \bibinfo{author}{\bibfnamefont{M.~S.} \bibnamefont{Dresselhaus}},
  \bibinfo{journal}{Phys. Rev. B} \textbf{\bibinfo{volume}{47}},
  \bibinfo{pages}{16631} (\bibinfo{year}{1993}{\natexlab{b}}).

\bibitem[{\citenamefont{Pichanusakorn and
  Prabhakar}(2010)}]{Pichanusakorn_nanostructured_thermoelectrics}
\bibinfo{author}{\bibfnamefont{P.}~\bibnamefont{Pichanusakorn}}
  \bibnamefont{and}
  \bibinfo{author}{\bibfnamefont{B.}~\bibnamefont{Prabhakar}},
  \bibinfo{journal}{Mater. Science and Engineering R}
  \textbf{\bibinfo{volume}{67}}, \bibinfo{pages}{19} (\bibinfo{year}{2010}).

\bibitem[{\citenamefont{Heremans}(2005)}]{Heremans_low_dimensional_thermoelectricity}
\bibinfo{author}{\bibfnamefont{J.~P.} \bibnamefont{Heremans}},
  \bibinfo{journal}{Proceedings of the XXXIV International School of
  Semiconducting Compounds} \textbf{\bibinfo{volume}{108}},
  \bibinfo{pages}{609} (\bibinfo{year}{2005}).

\bibitem[{\citenamefont{Dresselhaus et~al.}(2007)\citenamefont{Dresselhaus,
  Chen, Tang, Yang, Lee, Wang, Ren, Fleurial, and
  Gogna}}]{Dresselhaus_new_directions_low_dimensional}
\bibinfo{author}{\bibfnamefont{M.}~\bibnamefont{Dresselhaus}},
  \bibinfo{author}{\bibfnamefont{G.}~\bibnamefont{Chen}},
  \bibinfo{author}{\bibfnamefont{M.}~\bibnamefont{Tang}},
  \bibinfo{author}{\bibfnamefont{R.}~\bibnamefont{Yang}},
  \bibinfo{author}{\bibfnamefont{H.}~\bibnamefont{Lee}},
  \bibinfo{author}{\bibfnamefont{D.}~\bibnamefont{Wang}},
  \bibinfo{author}{\bibfnamefont{Z.}~\bibnamefont{Ren}},
  \bibinfo{author}{\bibfnamefont{J.-P.} \bibnamefont{Fleurial}},
  \bibnamefont{and} \bibinfo{author}{\bibfnamefont{P.}~\bibnamefont{Gogna}},
  \bibinfo{journal}{Adv. Mater.} \textbf{\bibinfo{volume}{19}},
  \bibinfo{pages}{1043} (\bibinfo{year}{2007}).

\bibitem[{\citenamefont{Hicks et~al.}(1996)\citenamefont{Hicks, Harman, Sun,
  and Dresselhaus}}]{Hicks_experimental_effect_quantum_well}
\bibinfo{author}{\bibfnamefont{L.~D.} \bibnamefont{Hicks}},
  \bibinfo{author}{\bibfnamefont{T.~C.} \bibnamefont{Harman}},
  \bibinfo{author}{\bibfnamefont{X.}~\bibnamefont{Sun}}, \bibnamefont{and}
  \bibinfo{author}{\bibfnamefont{M.~S.} \bibnamefont{Dresselhaus}},
  \bibinfo{journal}{Phys. Rev. B} \textbf{\bibinfo{volume}{53}},
  \bibinfo{pages}{10493} (\bibinfo{year}{1996}).

\bibitem[{\citenamefont{Venkatasubramanian
  et~al.}(2001)\citenamefont{Venkatasubramanian, Siivola, Colpitts, and
  O'Quinn}}]{Venkatasubramanian_thin_film}
\bibinfo{author}{\bibfnamefont{R.}~\bibnamefont{Venkatasubramanian}},
  \bibinfo{author}{\bibfnamefont{E.}~\bibnamefont{Siivola}},
  \bibinfo{author}{\bibfnamefont{T.}~\bibnamefont{Colpitts}}, \bibnamefont{and}
  \bibinfo{author}{\bibfnamefont{B.}~\bibnamefont{O'Quinn}},
  \bibinfo{journal}{Nature} \textbf{\bibinfo{volume}{413}},
  \bibinfo{pages}{597} (\bibinfo{year}{2001}).

\bibitem[{\citenamefont{Ohta et~al.}(2007)\citenamefont{Ohta, Kim, Mune,
  Mizoguchi, Nomura, Ohta, Nomura, Nakanishi, Ikuhara, Hirano
  et~al.}}]{Ohta_giant_seebeck_2D_electron_gas_STO}
\bibinfo{author}{\bibfnamefont{H.}~\bibnamefont{Ohta}},
  \bibinfo{author}{\bibfnamefont{S.}~\bibnamefont{Kim}},
  \bibinfo{author}{\bibfnamefont{Y.}~\bibnamefont{Mune}},
  \bibinfo{author}{\bibfnamefont{T.}~\bibnamefont{Mizoguchi}},
  \bibinfo{author}{\bibfnamefont{K.}~\bibnamefont{Nomura}},
  \bibinfo{author}{\bibfnamefont{S.}~\bibnamefont{Ohta}},
  \bibinfo{author}{\bibfnamefont{T.}~\bibnamefont{Nomura}},
  \bibinfo{author}{\bibfnamefont{Y.}~\bibnamefont{Nakanishi}},
  \bibinfo{author}{\bibfnamefont{Y.}~\bibnamefont{Ikuhara}},
  \bibinfo{author}{\bibfnamefont{M.}~\bibnamefont{Hirano}},
  \bibnamefont{et~al.}, \bibinfo{journal}{Nature Mater.}
  \textbf{\bibinfo{volume}{6}}, \bibinfo{pages}{129–134}
  (\bibinfo{year}{2007}).

\bibitem[{\citenamefont{Mahan and Sofo}(1996)}]{Mahan_Sofo_best_thermoelectric}
\bibinfo{author}{\bibfnamefont{G.~D.} \bibnamefont{Mahan}} \bibnamefont{and}
  \bibinfo{author}{\bibfnamefont{J.~O.} \bibnamefont{Sofo}},
  \bibinfo{journal}{Proc. Natl. Acad. Sci. USA} \textbf{\bibinfo{volume}{93}},
  \bibinfo{pages}{7436} (\bibinfo{year}{1996}).

\bibitem[{\citenamefont{Heremans et~al.}(2011)\citenamefont{Heremans,
  Wiendlocha, and Chamoire}}]{Heremans_resonant_levels_semiconductor}
\bibinfo{author}{\bibfnamefont{J.~P.} \bibnamefont{Heremans}},
  \bibinfo{author}{\bibfnamefont{B.}~\bibnamefont{Wiendlocha}},
  \bibnamefont{and} \bibinfo{author}{\bibfnamefont{A.~M.}
  \bibnamefont{Chamoire}}, \bibinfo{journal}{Energy Environ. Sci.}
  \textbf{\bibinfo{volume}{5}}, \bibinfo{pages}{5510} (\bibinfo{year}{2011}).

\bibitem[{\citenamefont{Heremans et~al.}(2008)\citenamefont{Heremans, Jovovic,
  Toberer, Saramat, Kurosaki, Charoenphakdee, Yamanaka, and
  Snyder}}]{Heremans_Enhancement_of_Thermoelectric_Efficiency_in_PbTe}
\bibinfo{author}{\bibfnamefont{J.~P.} \bibnamefont{Heremans}},
  \bibinfo{author}{\bibfnamefont{V.}~\bibnamefont{Jovovic}},
  \bibinfo{author}{\bibfnamefont{E.~S.} \bibnamefont{Toberer}},
  \bibinfo{author}{\bibfnamefont{A.}~\bibnamefont{Saramat}},
  \bibinfo{author}{\bibfnamefont{K.}~\bibnamefont{Kurosaki}},
  \bibinfo{author}{\bibfnamefont{A.}~\bibnamefont{Charoenphakdee}},
  \bibinfo{author}{\bibfnamefont{S.}~\bibnamefont{Yamanaka}}, \bibnamefont{and}
  \bibinfo{author}{\bibfnamefont{G.~J.} \bibnamefont{Snyder}},
  \bibinfo{journal}{Science} \textbf{\bibinfo{volume}{321}},
  \bibinfo{pages}{554} (\bibinfo{year}{2008}).

\bibitem[{\citenamefont{Nemov and
  Ravish}(2008)}]{Nemov_Ravish_review_thallium_lead_chalcogenides}
\bibinfo{author}{\bibfnamefont{S.~A.} \bibnamefont{Nemov}} \bibnamefont{and}
  \bibinfo{author}{\bibfnamefont{Y.~I.} \bibnamefont{Ravish}},
  \bibinfo{journal}{Phys.-Usp.} \textbf{\bibinfo{volume}{41}},
  \bibinfo{pages}{735} (\bibinfo{year}{2008}).

\bibitem[{\citenamefont{Jaworski et~al.}(2009)\citenamefont{Jaworski,
  Kulbachinskii, and Heremans}}]{Jaworski_Heremans_resonant_level_tin}
\bibinfo{author}{\bibfnamefont{C.~M.} \bibnamefont{Jaworski}},
  \bibinfo{author}{\bibfnamefont{V.}~\bibnamefont{Kulbachinskii}},
  \bibnamefont{and} \bibinfo{author}{\bibfnamefont{J.~P.}
  \bibnamefont{Heremans}}, \bibinfo{journal}{Phys. Rev. B}
  \textbf{\bibinfo{volume}{80}}, \bibinfo{pages}{233201}
  (\bibinfo{year}{2009}).

\bibitem[{\citenamefont{Zhang et~al.}(2013)\citenamefont{Zhang, Liao, Lan,
  Lukas, Liu, Esfarjani, Opeil, Broido, Chen, and
  Ren}}]{Zhang_resonant_Indium_SnTe}
\bibinfo{author}{\bibfnamefont{Q.}~\bibnamefont{Zhang}},
  \bibinfo{author}{\bibfnamefont{B.}~\bibnamefont{Liao}},
  \bibinfo{author}{\bibfnamefont{Y.}~\bibnamefont{Lan}},
  \bibinfo{author}{\bibfnamefont{K.}~\bibnamefont{Lukas}},
  \bibinfo{author}{\bibfnamefont{W.}~\bibnamefont{Liu}},
  \bibinfo{author}{\bibfnamefont{K.}~\bibnamefont{Esfarjani}},
  \bibinfo{author}{\bibfnamefont{C.}~\bibnamefont{Opeil}},
  \bibinfo{author}{\bibfnamefont{D.}~\bibnamefont{Broido}},
  \bibinfo{author}{\bibfnamefont{G.}~\bibnamefont{Chen}}, \bibnamefont{and}
  \bibinfo{author}{\bibfnamefont{Z.}~\bibnamefont{Ren}},
  \bibinfo{journal}{PNAS} \textbf{\bibinfo{volume}{110}},
  \bibinfo{pages}{13261} (\bibinfo{year}{2013}).

\bibitem[{\citenamefont{Tan et~al.}(2015)\citenamefont{Tan, Zeier, Shi, Wang,
  Snyder, Dravid, and Kanatzidis}}]{Tan_high_performance_SnTe_resonant_levels}
\bibinfo{author}{\bibfnamefont{G.}~\bibnamefont{Tan}},
  \bibinfo{author}{\bibfnamefont{W.~G.} \bibnamefont{Zeier}},
  \bibinfo{author}{\bibfnamefont{F.}~\bibnamefont{Shi}},
  \bibinfo{author}{\bibfnamefont{P.}~\bibnamefont{Wang}},
  \bibinfo{author}{\bibfnamefont{G.~J.} \bibnamefont{Snyder}},
  \bibinfo{author}{\bibfnamefont{V.~P.} \bibnamefont{Dravid}},
  \bibnamefont{and} \bibinfo{author}{\bibfnamefont{M.~G.}
  \bibnamefont{Kanatzidis}}, \bibinfo{journal}{Chem. Mater.}
  \textbf{\bibinfo{volume}{27}}, \bibinfo{pages}{7801} (\bibinfo{year}{2015}).

\bibitem[{\citenamefont{Zhang et~al.}(2012)\citenamefont{Zhang, Wang, Liu,
  Wang, Yu, Zhang, Tian, Ni, Lee, Esfarjani
  et~al.}}]{Zhang_enhancement_resonant_states_PbSe}
\bibinfo{author}{\bibfnamefont{Q.}~\bibnamefont{Zhang}},
  \bibinfo{author}{\bibfnamefont{H.}~\bibnamefont{Wang}},
  \bibinfo{author}{\bibfnamefont{W.}~\bibnamefont{Liu}},
  \bibinfo{author}{\bibfnamefont{H.}~\bibnamefont{Wang}},
  \bibinfo{author}{\bibfnamefont{B.}~\bibnamefont{Yu}},
  \bibinfo{author}{\bibfnamefont{Q.}~\bibnamefont{Zhang}},
  \bibinfo{author}{\bibfnamefont{Z.}~\bibnamefont{Tian}},
  \bibinfo{author}{\bibfnamefont{G.}~\bibnamefont{Ni}},
  \bibinfo{author}{\bibfnamefont{S.}~\bibnamefont{Lee}},
  \bibinfo{author}{\bibfnamefont{K.}~\bibnamefont{Esfarjani}},
  \bibnamefont{et~al.}, \bibinfo{journal}{Energy Environ. Sci.}
  \textbf{\bibinfo{volume}{5}}, \bibinfo{pages}{5246} (\bibinfo{year}{2012}).

\bibitem[{\citenamefont{Li et~al.}(2014)\citenamefont{Li, Liu, Li, Song, Liu,
  and Ao}}]{Li_Sn_substitution_YbAl3}
\bibinfo{author}{\bibfnamefont{J.}~\bibnamefont{Li}},
  \bibinfo{author}{\bibfnamefont{X.}~\bibnamefont{Liu}},
  \bibinfo{author}{\bibfnamefont{Y.}~\bibnamefont{Li}},
  \bibinfo{author}{\bibfnamefont{S.}~\bibnamefont{Song}},
  \bibinfo{author}{\bibfnamefont{F.}~\bibnamefont{Liu}}, \bibnamefont{and}
  \bibinfo{author}{\bibfnamefont{W.}~\bibnamefont{Ao}}, \bibinfo{journal}{J. of
  Alloys and Compounds} \textbf{\bibinfo{volume}{600}}, \bibinfo{pages}{8}
  (\bibinfo{year}{2014}).

\bibitem[{\citenamefont{Rowe et~al.}(2002)\citenamefont{Rowe, Kuznetsov,
  Kuznetsova, and Min}}]{Rowe_transport_properties_YbAl3}
\bibinfo{author}{\bibfnamefont{D.}~\bibnamefont{Rowe}},
  \bibinfo{author}{\bibfnamefont{V.~L.} \bibnamefont{Kuznetsov}},
  \bibinfo{author}{\bibfnamefont{L.~A.} \bibnamefont{Kuznetsova}},
  \bibnamefont{and} \bibinfo{author}{\bibfnamefont{G.}~\bibnamefont{Min}},
  \bibinfo{journal}{J. Phys. D: Appl. Phys.} \textbf{\bibinfo{volume}{35}},
  \bibinfo{pages}{2183–2186} (\bibinfo{year}{2002}).

\bibitem[{\citenamefont{Zhou et~al.}(2009)\citenamefont{Zhou, Sun, Cheng, and
  Zhang}}]{Zhou_electronic_structure_YbAl3}
\bibinfo{author}{\bibfnamefont{J.}~\bibnamefont{Zhou}},
  \bibinfo{author}{\bibfnamefont{Z.}~\bibnamefont{Sun}},
  \bibinfo{author}{\bibfnamefont{X.}~\bibnamefont{Cheng}}, \bibnamefont{and}
  \bibinfo{author}{\bibfnamefont{Y.}~\bibnamefont{Zhang}},
  \bibinfo{journal}{Intermetallics} \textbf{\bibinfo{volume}{17}},
  \bibinfo{pages}{995–999} (\bibinfo{year}{2009}).

\bibitem[{\citenamefont{Yang}(2014)}]{Yang_enhancing_thermoelectric_doping_electronegativites_distinct}
\bibinfo{author}{\bibfnamefont{X.~H.} \bibnamefont{Yang}}, \bibinfo{journal}{J.
  Alloys. Compounds.} \textbf{\bibinfo{volume}{594}}, \bibinfo{pages}{70}
  (\bibinfo{year}{2014}).

\bibitem[{\citenamefont{Singh}(2010)}]{Singh_thermopower_PbTe_boltzmann}
\bibinfo{author}{\bibfnamefont{D.~J.} \bibnamefont{Singh}},
  \bibinfo{journal}{Phys. Rev. B} \textbf{\bibinfo{volume}{81}},
  \bibinfo{pages}{195217} (\bibinfo{year}{2010}).

\bibitem[{\citenamefont{Peng et~al.}(2011)\citenamefont{Peng, Song, Kanatzidis,
  and Freeman}}]{Peng_electronic_structure_transport_properties_doped_PbSe}
\bibinfo{author}{\bibfnamefont{H.}~\bibnamefont{Peng}},
  \bibinfo{author}{\bibfnamefont{J.-H.} \bibnamefont{Song}},
  \bibinfo{author}{\bibfnamefont{M.~G.} \bibnamefont{Kanatzidis}},
  \bibnamefont{and} \bibinfo{author}{\bibfnamefont{A.~J.}
  \bibnamefont{Freeman}}, \bibinfo{journal}{Phys. Rev. B}
  \textbf{\bibinfo{volume}{84}}, \bibinfo{pages}{125207}
  (\bibinfo{year}{2011}).

\bibitem[{\citenamefont{Parker and
  Singh}(2010)}]{Parker_Singh_high_temperature_dependance_heavily_doped_PbSe}
\bibinfo{author}{\bibfnamefont{D.}~\bibnamefont{Parker}} \bibnamefont{and}
  \bibinfo{author}{\bibfnamefont{D.~J.} \bibnamefont{Singh}},
  \bibinfo{journal}{Phys. Rev. B} \textbf{\bibinfo{volume}{82}},
  \bibinfo{pages}{035204} (\bibinfo{year}{2010}).

\bibitem[{\citenamefont{Yang et~al.}(2015{\natexlab{a}})\citenamefont{Yang,
  Qin, Li, Zhang, Song, Liu, Wang, and
  Xin}}]{Yang_electronic_properties_CdTe_doped}
\bibinfo{author}{\bibfnamefont{X.}~\bibnamefont{Yang}},
  \bibinfo{author}{\bibfnamefont{X.}~\bibnamefont{Qin}},
  \bibinfo{author}{\bibfnamefont{D.}~\bibnamefont{Li}},
  \bibinfo{author}{\bibfnamefont{J.}~\bibnamefont{Zhang}},
  \bibinfo{author}{\bibfnamefont{C.}~\bibnamefont{Song}},
  \bibinfo{author}{\bibfnamefont{Y.}~\bibnamefont{Liu}},
  \bibinfo{author}{\bibfnamefont{L.}~\bibnamefont{Wang}}, \bibnamefont{and}
  \bibinfo{author}{\bibfnamefont{H.}~\bibnamefont{Xin}}, \bibinfo{journal}{J.
  of Physics and Chem. of Solids} \textbf{\bibinfo{volume}{86}},
  \bibinfo{pages}{74} (\bibinfo{year}{2015}{\natexlab{a}}).

\bibitem[{\citenamefont{Ashcroft and Mermin}(1976)}]{ashcroft}
\bibinfo{author}{\bibfnamefont{N.~W.} \bibnamefont{Ashcroft}} \bibnamefont{and}
  \bibinfo{author}{\bibfnamefont{N.~D.} \bibnamefont{Mermin}},
  \emph{\bibinfo{title}{Solid State Physics}} (\bibinfo{publisher}{Saunders
  College Publishing}, \bibinfo{year}{1976}).

\bibitem[{\citenamefont{Lee et~al.}(2010)\citenamefont{Lee, Wu, and
  Grossman}}]{Lee_enhancing_thermoelectric_mismatched_doping}
\bibinfo{author}{\bibfnamefont{J.-H.} \bibnamefont{Lee}},
  \bibinfo{author}{\bibfnamefont{J.}~\bibnamefont{Wu}}, \bibnamefont{and}
  \bibinfo{author}{\bibfnamefont{J.~C.} \bibnamefont{Grossman}},
  \bibinfo{journal}{Phys. Rev. Lett.} \textbf{\bibinfo{volume}{104}},
  \bibinfo{pages}{016602} (\bibinfo{year}{2010}).

\bibitem[{\citenamefont{Yang et~al.}(2015{\natexlab{b}})\citenamefont{Yang, Xi,
  Qiu, Wu, Shi, Chen, Yang, Zhang, Uher, and
  Singh}}]{Yang_theory_experiment_perspective}
\bibinfo{author}{\bibfnamefont{J.}~\bibnamefont{Yang}},
  \bibinfo{author}{\bibfnamefont{L.}~\bibnamefont{Xi}},
  \bibinfo{author}{\bibfnamefont{W.}~\bibnamefont{Qiu}},
  \bibinfo{author}{\bibfnamefont{L.}~\bibnamefont{Wu}},
  \bibinfo{author}{\bibfnamefont{X.}~\bibnamefont{Shi}},
  \bibinfo{author}{\bibfnamefont{L.}~\bibnamefont{Chen}},
  \bibinfo{author}{\bibfnamefont{J.}~\bibnamefont{Yang}},
  \bibinfo{author}{\bibfnamefont{W.}~\bibnamefont{Zhang}},
  \bibinfo{author}{\bibfnamefont{C.}~\bibnamefont{Uher}}, \bibnamefont{and}
  \bibinfo{author}{\bibfnamefont{D.~J.} \bibnamefont{Singh}},
  \bibinfo{journal}{npj Computational Materials} \textbf{\bibinfo{volume}{2}},
  \bibinfo{pages}{15015} (\bibinfo{year}{2015}{\natexlab{b}}).

\bibitem[{\citenamefont{Zhou et~al.}(2011)\citenamefont{Zhou, Yang, Chen, and
  Dresselhaus}}]{Zhou_Dresselhaus_optimal_bandwidth_efficiency_thermoelectrics}
\bibinfo{author}{\bibfnamefont{J.}~\bibnamefont{Zhou}},
  \bibinfo{author}{\bibfnamefont{R.}~\bibnamefont{Yang}},
  \bibinfo{author}{\bibfnamefont{G.}~\bibnamefont{Chen}}, \bibnamefont{and}
  \bibinfo{author}{\bibfnamefont{M.~S.} \bibnamefont{Dresselhaus}},
  \bibinfo{journal}{Phys. Rev. Lett.} \textbf{\bibinfo{volume}{107}},
  \bibinfo{pages}{226601} (\bibinfo{year}{2011}).

\bibitem[{\citenamefont{Jeong and
  Lundstrom}(2012)}]{Jeong_best_bandstructure_thermoelectric_landauer}
\bibinfo{author}{\bibfnamefont{C.}~\bibnamefont{Jeong}} \bibnamefont{and}
  \bibinfo{author}{\bibfnamefont{M.~S.} \bibnamefont{Lundstrom}},
  \bibinfo{journal}{J. Appl. Physics.} \textbf{\bibinfo{volume}{111}},
  \bibinfo{pages}{113707} (\bibinfo{year}{2012}).

\bibitem[{\citenamefont{Bilc et~al.}(2015)\citenamefont{Bilc, Hautier,
  Waroquiers, Rignanese, and
  Ghosez}}]{Bilc_low_dimensional_band_engineering_highly_directional}
\bibinfo{author}{\bibfnamefont{D.~I.} \bibnamefont{Bilc}},
  \bibinfo{author}{\bibfnamefont{G.}~\bibnamefont{Hautier}},
  \bibinfo{author}{\bibfnamefont{D.}~\bibnamefont{Waroquiers}},
  \bibinfo{author}{\bibfnamefont{G.-M.} \bibnamefont{Rignanese}},
  \bibnamefont{and} \bibinfo{author}{\bibfnamefont{P.}~\bibnamefont{Ghosez}},
  \bibinfo{journal}{Phys. Rev. Lett.} \textbf{\bibinfo{volume}{114}},
  \bibinfo{pages}{136601} (\bibinfo{year}{2015}).

\bibitem[{\citenamefont{Bilc et~al.}(2016)\citenamefont{Bilc, Floare, Zârbo,
  Garabagiu, Lemal, and Ghosez}}]{Bilc_first_principles_modeling_sto}
\bibinfo{author}{\bibfnamefont{D.~I.} \bibnamefont{Bilc}},
  \bibinfo{author}{\bibfnamefont{C.~G.} \bibnamefont{Floare}},
  \bibinfo{author}{\bibfnamefont{L.~P.} \bibnamefont{Zârbo}},
  \bibinfo{author}{\bibfnamefont{S.}~\bibnamefont{Garabagiu}},
  \bibinfo{author}{\bibfnamefont{S.}~\bibnamefont{Lemal}}, \bibnamefont{and}
  \bibinfo{author}{\bibfnamefont{P.}~\bibnamefont{Ghosez}},
  \bibinfo{journal}{J. Phys. Chem. C} \textbf{\bibinfo{volume}{120}},
  \bibinfo{pages}{25678} (\bibinfo{year}{2016}).

\bibitem[{\citenamefont{Pei et~al.}(2012)\citenamefont{Pei, Wang, and
  J.}}]{Pei_band_engineering}
\bibinfo{author}{\bibfnamefont{Y.}~\bibnamefont{Pei}},
  \bibinfo{author}{\bibfnamefont{H.}~\bibnamefont{Wang}}, \bibnamefont{and}
  \bibinfo{author}{\bibfnamefont{S.~G.} \bibnamefont{J.}},
  \bibinfo{journal}{Adv. Mater.} \textbf{\bibinfo{volume}{24}},
  \bibinfo{pages}{6125} (\bibinfo{year}{2012}).

\bibitem[{\citenamefont{Pei et~al.}(2011)\citenamefont{Pei, Shi, LaLonde, Wang,
  Chen, and Snyder}}]{Pei_Snyder_convergence_electronic_bands_high_performance}
\bibinfo{author}{\bibfnamefont{Y.}~\bibnamefont{Pei}},
  \bibinfo{author}{\bibfnamefont{X.}~\bibnamefont{Shi}},
  \bibinfo{author}{\bibfnamefont{A.}~\bibnamefont{LaLonde}},
  \bibinfo{author}{\bibfnamefont{H.}~\bibnamefont{Wang}},
  \bibinfo{author}{\bibfnamefont{L.}~\bibnamefont{Chen}}, \bibnamefont{and}
  \bibinfo{author}{\bibfnamefont{G.~J.} \bibnamefont{Snyder}},
  \bibinfo{journal}{Nature} \textbf{\bibinfo{volume}{473}}, \bibinfo{pages}{66}
  (\bibinfo{year}{2011}).

\bibitem[{\citenamefont{Zhong et~al.}(2013)\citenamefont{Zhong, Zhang, and
  Held}}]{Zhong_Held_quantum_confinement_oxide_tight_binding}
\bibinfo{author}{\bibfnamefont{Z.}~\bibnamefont{Zhong}},
  \bibinfo{author}{\bibfnamefont{Q.}~\bibnamefont{Zhang}}, \bibnamefont{and}
  \bibinfo{author}{\bibfnamefont{K.}~\bibnamefont{Held}},
  \bibinfo{journal}{Phys. Rev. B} \textbf{\bibinfo{volume}{88}},
  \bibinfo{pages}{125401} (\bibinfo{year}{2013}).

\bibitem[{\citenamefont{Zhong et~al.}(2012)\citenamefont{Zhong, Wissgott, Held,
  and
  Sangiovanni}}]{Zhong_Held_microscopic_understanding_orbital_splitting_oxide_interfaces}
\bibinfo{author}{\bibfnamefont{Z.}~\bibnamefont{Zhong}},
  \bibinfo{author}{\bibfnamefont{P.}~\bibnamefont{Wissgott}},
  \bibinfo{author}{\bibfnamefont{K.}~\bibnamefont{Held}}, \bibnamefont{and}
  \bibinfo{author}{\bibfnamefont{G.}~\bibnamefont{Sangiovanni}},
  \bibinfo{journal}{EPL} \textbf{\bibinfo{volume}{99}}, \bibinfo{pages}{37011}
  (\bibinfo{year}{2012}).

\bibitem[{\citenamefont{{Bouzerar} et~al.}(2017)\citenamefont{{Bouzerar},
  {Th{\'e}baud}, {Adessi}, {Debord}, {Apreutesei}, {Bachelet}, and
  {Pailh{\`e}s}}}]{Bouzerar_unified_modelling_thermoelectric_properties_STO}
\bibinfo{author}{\bibfnamefont{G.}~\bibnamefont{{Bouzerar}}},
  \bibinfo{author}{\bibfnamefont{S.}~\bibnamefont{{Th{\'e}baud}}},
  \bibinfo{author}{\bibfnamefont{C.}~\bibnamefont{{Adessi}}},
  \bibinfo{author}{\bibfnamefont{R.}~\bibnamefont{{Debord}}},
  \bibinfo{author}{\bibfnamefont{M.}~\bibnamefont{{Apreutesei}}},
  \bibinfo{author}{\bibfnamefont{R.}~\bibnamefont{{Bachelet}}},
  \bibnamefont{and}
  \bibinfo{author}{\bibfnamefont{S.}~\bibnamefont{{Pailh{\`e}s}}},
  \bibinfo{journal}{arXiv:1702.0275 [cond-mat.mtrl-sci]}
  (\bibinfo{year}{2017}).

\bibitem[{\citenamefont{Scalapino et~al.}(1993)\citenamefont{Scalapino, White,
  and Zhang}}]{Scalapino_insulator_metal_criteria}
\bibinfo{author}{\bibfnamefont{D.~J.} \bibnamefont{Scalapino}},
  \bibinfo{author}{\bibfnamefont{S.~R.} \bibnamefont{White}}, \bibnamefont{and}
  \bibinfo{author}{\bibfnamefont{S.}~\bibnamefont{Zhang}},
  \bibinfo{journal}{Phys. Rev. B} \textbf{\bibinfo{volume}{47}},
  \bibinfo{pages}{7995} (\bibinfo{year}{1993}).

\bibitem[{\citenamefont{Kohn}(1964)}]{Kohn_theory_insulating_state}
\bibinfo{author}{\bibfnamefont{W.}~\bibnamefont{Kohn}}, \bibinfo{journal}{Phys.
  Rev. A} \textbf{\bibinfo{volume}{133}}, \bibinfo{pages}{171}
  (\bibinfo{year}{1964}).

\bibitem[{\citenamefont{Millis and
  Coppersmith}(1990)}]{Millis_Coppersmith_optical_spectral_weight}
\bibinfo{author}{\bibfnamefont{A.~J.} \bibnamefont{Millis}} \bibnamefont{and}
  \bibinfo{author}{\bibfnamefont{S.~N.} \bibnamefont{Coppersmith}},
  \bibinfo{journal}{Phys. Rev. B} \textbf{\bibinfo{volume}{42}},
  \bibinfo{pages}{10 807} (\bibinfo{year}{1990}).

\bibitem[{\citenamefont{Bouzerar and
  Bouzerar}(2011)}]{Bouzerar_optical_conductivity}
\bibinfo{author}{\bibfnamefont{G.}~\bibnamefont{Bouzerar}} \bibnamefont{and}
  \bibinfo{author}{\bibfnamefont{R.}~\bibnamefont{Bouzerar}},
  \bibinfo{journal}{New J. Phys.} \textbf{\bibinfo{volume}{13}},
  \bibinfo{pages}{023002} (\bibinfo{year}{2011}).

\end{thebibliography}
\end{document}